 \documentclass[twocolumn,preprint2,trackchanges]{aastex62}
\hypersetup{linkcolor=red,citecolor=blue,filecolor=cyan,urlcolor=magenta}
\usepackage{aas_macros}
\usepackage{grffile}
\usepackage{amsmath}
\usepackage{amsfonts}
\usepackage{bm}
\graphicspath{{./}{figures/}}

\received{---}
\revised{---}
\accepted{---}
\submitjournal{ApJ}

\shorttitle{A fast and exact $w$-stacking $w$-projection hybrid}
\shortauthors{Pratley et al.}
\begin{document}
\title{A fast and exact $w$-stacking and $w$-projection hybrid algorithm for wide-field interferometric imaging}

\correspondingauthor{Luke Pratley}
\email{Luke.Pratley@gmail.com}

\author{Luke Pratley}
\affiliation{Mullard Space Science Laboratory (MSSL),\\ 
University College London (UCL),\\ 
Holmbury St Mary, Surrey RH5 6NT, UK}

\author{Melanie Johnston-Hollitt}
\affiliation{International Centre for Radio Astronomy Research (ICRAR), \\
Curtin University,\\
1 Turner Ave., Technology Park, Bentley, 6102,	WA, Australia}

\author{Jason D. McEwen}
\affiliation{Mullard Space Science Laboratory (MSSL),\\ 
University College London (UCL),\\ 
Holmbury St Mary, Surrey RH5 6NT, UK}



\begin{abstract}
The standard wide-field imaging technique, the $w$-projection, allows correction for wide-fields of view for non-coplanar radio interferometric arrays. However, calculating exact corrections for each measurement has not been possible due to the amount of computation required at high resolution and with the large number of visibilities from current interferometers. The required accuracy and computational cost of these corrections is one of the largest unsolved challenges facing next generation radio interferometers such as the Square Kilometre Array. We show that the same calculation can be performed with a radially symmetric $w$-projection kernel, where we use one dimensional adaptive quadrature to calculate the resulting Hankel transform, decreasing the computation required for kernel generation by several orders of magnitude, whilst preserving the accuracy. We confirm that the radial $w$-projection kernel is accurate to approximately 1\% by imaging the zero-spacing with an added $w$-term. We demonstrate the potential of our radially symmetric $w$-projection kernel via sparse image reconstruction, using the software package PURIFY. {We develop a distributed $w$-stacking and $w$-projection hybrid algorithm. We apply this algorithm to individually correct for non-coplanar effects in 17.5 million visibilities over a $25$ by $25$ degree field of view MWA observation for image reconstruction. Such a level of accuracy and scalability is not possible with standard $w$-projection kernel generation methods. This demonstrates that we can scale to a large number of measurements with large image sizes whilst still maintaining both speed and accuracy. }
\end{abstract}

% look up list of allowable keywords for this section

\section{INTRODUCTION}
\label{sec:intro}
Since the advent of radio interferometry in the 1940s \citep{Pawsey46,Ryle48} radio astronomers have built an impressive suite of interferometric imaging techniques to allow signals from collections of antennas to be used collectively to image astronomical sources. As successive generations of interferometric arrays were built and operated, techniques were developed to obtain an estimate of the true sky brightness distribution, and to correct for different instrumental affects inherent in the process. 
Among these methods are processes such as deconvolution of the antenna response, so-called `CLEANing' \citep{hog74,sch78,ste84,Us16}, and methods to account for wide-field and other direction dependent effects (DDEs) such as $w$-projection  \citep{cor08b} and $a$-projection \citep{Bhat08}. 

In the past where the field of view of instruments was relatively small, it was common practice to assume curvature was negligible and proceed with a two dimensional Fourier transform over the $uv$-plane (using cartesian coordinates). With the arrival of next generation telescopes, such as the LOw Frequency ARray (LOFAR; \citealp{vH13}), Murchison Widefield Array (MWA; \citealp{tin13}), and Hydrogen Epoch of Reionization Array (HERA; \citealp{par17}), telescopes became non-coplanar arrays with extremely large fields of view. Such instruments are precursors to the low frequency component of the Square Kilometre Array (SKA-LOW), and are already encountering `big data' challenges. Imaging and correcting for DDEs (with wide-field of view DDEs being the most basic) are among the most computationally intensive and critical challenges that needs to be solved if the SKA is to meet its scientific goals, in areas such as the Epoch of Reionization (EoR) \citep{koo15} and Cosmic Magnetism \citep{mjh15}. Until now, the approach to account for the third Fourier dimension, $w$, has been to use mathematical approximations to correct for this term and the associated wide-field effects in the measurement equation, reducing the problem back to a two dimensional Fourier transform via the so-called `$w$-projection algorithm' \citep{cor08b, tas13, off14}. 

However, the $w$-projection algorithm kernels, used to correct for non-coplanar array and sky curvature, to date have been computationally expensive to calculate, with kernel generation dominated by the Fast Fourier Transform (FFT) \citep{sca15}. In particular the gridding kernel (anti-aliasing kernel) and $w$-chirp are multiplied in image space, and then an FFT is applied to generate the $w$-projection kernel \citep{corn11}. This means it has not been possible to generate a kernel for each $w$-term individually, instead they are generated as $w$-planes, approximately correcting for a group of $w$-terms. 

For extremely wide-fields of view, this becomes expensive in computation and memory, and requires both high resolution sampling to model the spherical curvature and extra zero padding to increase sub-pixel accuracy in the $uv$-domain. Such a cost in kernel construction has motivated alternative imaging strategies, such as image domain gridding \citep{van18}. 
Even for small fields of view with high resolution, it is not possible to perform an FFT for each visibility on large data sets, limiting the kernel calculation to a small number of $w$-planes. However, \citet{mer16} mathematically showed that for narrow fields of view the $w$-projection kernel can be approximated as separable into a product of two 1d kernels, reducing the resources required to generate $w$-planes.

In this work, we set out to improve the analytic understanding of wide-field interferometry, in the hopes that it would provide clues on how to improve the strategy of expensive kernel construction. 
We start by presenting the non-standard analytic expression for the 3d Fourier transform used to create the $w$-projection kernel. Then using the analytic expression for the Fourier transform of a spherical shell and enforcing the horizon window with a convolution kernel, we arrive at the 3d expression for the sky curvature and horizon in the $uvw$-domain. The real component of the kernel is a radial Sinc function in $uvw$. It is also clear that the horizon window produces the imaginary component, which is a Hilbert transform of the real component.
With this understanding, we investigate construction through 3d convolution in the $uvw$-domain to generate gridding kernels. However, this proves computationally challenging due to rapid osculations and large function support\footnote{By the support of a function we mean the region of the domain where the function has non-zero output.}.

We find it is less challenging to generate the $w$-projection kernel via a Fourier integral using 2d adaptive quadrature, due to the smoothness of the window function and the chirp. However, under the condition that the window function has radial symmetry, this 2d Fourier integral is equivalent to 1d Hankel transform. We show that such a 1d Hankel transform can be fast and accurately computed with adaptive quadrature compared to the 2d Fourier integral, and produces the same imaging results. 

We discuss the computational impact of having a 1d radially symmetric $w$-projection kernel, such as reducing the dimension of $w$-planes from 2d to 1d radial planes, allowing new possibilities for reducing kernel construction costs.

Lastly, we provide a demonstration of exact correction of the $w$-component to simulated big data sparse image reconstruction using the software package PURIFY \citep{car14,LP18}, using the hybrid of $w$-stacking and $w$-projection with distributed computation on a high performance computing cluster. {Correction of the $w$-component for each measurement is only possible with the developments in this work, a radially symmetric $w$-projection kernel and distributed computation with $w$-stacking.}

The developments presented here provide an accurate route for reducing the computational overhead for next generation wide-field imaging, thus providing a step forward on the path to realizing the SKA. 

This work starts with an introduction to the interferometric measurement equation and the $w$-projection algorithm in Section \ref{sec:radio_measurement_equation}, Section \ref{sec:3dproj} extends the $w$-projection derivation starting from a 3d setting. The calculation of a 1d radially symmetric $w$-projection kernel is derived in Section \ref{sec:quadrature}. The 1d radially symmetric kernel is then numerically validated and benchmarked in Section \ref{sec:numerical_tests}. Section \ref{sec:MPI_algo} details and demonstrates the computationally distributed $w$-stacking and $w$-projection hybrid algorithm that is possible with a 1d $w$-projection kernel. This work is concluded in Section \ref{sec:conclusion}.
																						
\section{Interferometric measurement equation}
\label{sec:radio_measurement_equation}
The interferometric measurement equation for a radio telescope can be represented by the following integral
\begin{equation}
\begin{split}
	y(u, v, w^\prime) = \int x(l, m) a(l, m)\frac{{\rm e}^{-2\pi i w^\prime(\sqrt{1 -l^2 - m^2} - 1)}}{\sqrt{1 -l^2 - m^2}}\\
	\times{\rm e}^{-2\pi i (lu + mv)}\,  {\rm d}l{\rm d}m\, ,
	\end{split}
	\label{eq:meas_eq}
\end{equation}
$(u, v, w^\prime)$ are the baseline coordinates and $(l, m, n)$ are directional cosines restricted to the unit sphere. In this work, we define $w^\prime = w + \bar{w}$, where $\bar{w}$ is the average value of $w$-terms, and $w$ is the effective $w$-component (with zero mean).
$x$ is the sky brightness, $n(\bm{l})= \sqrt{1 -l^2-m^2}$ is a parametrization of the upper hemisphere, and $a$ includes direction dependent effects such as the primary beam and Field of View (FoV). The measurement equation is a mathematical model of the measurement operation that allows one to calculate model measurements $y$ when provided with a sky model $x$. Having such a measurement equation allows one to find a best fit model of the sky brightness, for a given set of (incomplete) measurements. Many techniques are available for inverting a measurement equation in an attempt to find a best fit model. This includes traditional methods such as CLEAN \citep{hog74} and Maximum Entropy \citep{Ables74,Cornwell85}, and state of the art deconvolution methods such as Sparse Regularization algorithms \citep{ono16,LP18,da18}. There are many other variations of the measurement equation, that can include general direction dependent effects and polarization \citep{mce08, smi11, pri15}. But, all interferometric measurement equations can be derived from the van Cittert-Zernike theorem \citep{zer38}.
																						
This measurement equation is typically approximated by a non-uniform fast Fourier transform, since it reduces the computational complexity from $\mathcal{O}(MN)$ to $\mathcal{O}(MJ^2 + N\log N)$, where $N$ is the number of pixels $M$ is the number of visibilities, and $J$ is the number of weights to interpolate off the fast Fourier transform (FFT) grid for each axis \citep{fes03,tho08}. This process is traditionally known as degridding. The version of the measurement equation relevant in this work is represented by the following linear operations 
\begin{equation}
	\bm{y} = \bm{\mathsf{W}}\bm{\mathsf{G}}\bm{\mathsf{C}}\bm{\mathsf{F}}\bm{\mathsf{Z}}\bm{\mathsf{S}} \bm{x}\, 
	\label{eq:matrix_equation}
\end{equation}
$\bm{\mathsf{S}}$ represents a gridding correction and correction of baseline independent effects such as $\bar{w}$, $\bm{\mathsf{Z}}$ represents zero padding of the image, $\bm{\mathsf{F}}$ is an FFT, $\bm{\mathsf{G}}$ represents a sparse circular convolution matrix that interpolates measurements off the grid and the combined $\bm{\mathsf{G}}\bm{\mathsf{C}}$ includes baseline dependent effects such as variations in the primary beam and $w$-component in the interpolation, and $\bm{\mathsf{W}}$ are weights applied to the measurements. This linear operator represents the application of the measurement equation, so is typically called a measurement operator $\bm{\mathsf{\Phi}} = \bm{\mathsf{W}}\bm{\mathsf{G}}\bm{\mathsf{C}}\bm{\mathsf{F}}\bm{\mathsf{Z}}\bm{\mathsf{S}}$ with $\bm{\mathsf{\Phi}} \in \mathbb{C}^{M \times N}$. 
																						
In this case, $\bm{x}_i = x(\bm{l}_i)$ and $\bm{y}_i = y(\bm{u}_i)$ are discrete vectors in $\mathbb{C}^{N \times 1}$ and $\mathbb{C}^{M \times 1}$ of the sky brightness and visibilities, respectively.
																						
Since the measurement operator is linear it has an adjoint operator $\bm{\mathsf{\Phi}}^\dagger$, which essentially, consists of applying these operators in reverse. Additionally, it is possible to represent these operators in matrix form, however, this is not always efficient or practical.
																						
The dirty map can be calculated by $\bm{\mathsf{\Phi}}^\dagger\bm{y}$, and the residuals by $\bm{\mathsf{\Phi}}^\dagger\bm{\mathsf{\Phi}}\bm{x} - \bm{\mathsf{\Phi}}^\dagger\bm{y}$.

\subsection{Gridding and degridding}
Degridding, also known as the NUFFT, is the process of applying the linear operators $\bm{\mathsf{G}}\bm{\mathsf{F}}\bm{\mathsf{Z}}\bm{\mathsf{S}}$. There are many works in the literature describing this process (see Section 4 of \citealp{LP18} for a brief review). The zero padding, $\bm{\mathsf{Z}}$, (normally by a factor of 2) is to increase accuracy of degridding/gridding of visibilities, by up sampling in the Fourier domain. The choice of interpolation weights in $\bm{\mathsf{G}}$, known as the gridding kernel, affects the aliasing error, where ghost periodic structures can appear in the dirty map from outside the imaged region. An ideal gridding kernel would be a sinc interpolation kernel, which would prevent any ghosting from the imaged region with a box function, but this has a large support (highly non localized). Well known kernels, such as prolate spheroidal wave functions (PSWF) and Kaiser Bessel functions, are known to suppress the ghosting through apodisation while having minimal support on the Fourier grid \citep{fes03,off14,LP18}. This apoidisation is then corrected for with the gridding correction $\bm{\mathsf{S}}$.
																						
Importantly, the size of the cell in a grid is inversely proportional to the field of view, and the number of cells in a grid determines the resolution of the image.

\subsection{The projection algorithm}
The projection algorithm has been developed to model baseline dependent effects. Typically, DDEs in the measurement equation such as the primary beam and $w$-term are multiplied with the sky intensity in the image domain. Since they are baseline dependent, a separate primary beam and $w$-term would need to be multiplied for each baseline -- which is computationally inefficient as this involves applying a different gridding/degridding process for each baseline.
																						
If we define our baseline dependent DDEs as
\begin{equation}
	\begin{split}
		c(l, m; w) = a(l, m)\frac{{\rm e}^{-2\pi i w(\sqrt{1 - l^2 - m^2} - 1)}}{\sqrt{1 - l^2 - m^2}}\, , 
	\end{split}
\end{equation}
the measurement equation can be expressed as
\begin{equation}
	\begin{split}
		y(u, v, \bar{w} + w) = \int x(l, m){\rm e}^{-2\pi i \bar{w}(\sqrt{1 - l^2 - m^2} - 1)}\quad\quad\quad\quad\quad\quad \\
		\times c(l, m; w){\rm e}^{-2\pi i (lu + mv)}\,  {\rm d}l{\rm d}m \, .
	\end{split}
\end{equation}
We can use the convolution theorem, which states that for functions $f$ and $g$ we have  $\mathcal{F}^{-1}\{\mathcal{F}\{f\}\mathcal{F}\{g\}\} = f \star g$, where convolution in 3d is defined as
\begin{equation}
	\begin{split}
		(f \star g)(x, y, z) = \int_{-\infty}^{+\infty}\int_{-\infty}^{+\infty}\int_{-\infty}^{+\infty} f(t, r, q)\quad\quad\quad\quad\quad\quad \\
		\times g(x - t, y - r, z - q)\,{\rm d}t{\rm d}r{\rm d}q\, .
	\end{split}
\end{equation}
This produces the expression
\begin{equation}
	y(u, v, w) = \tilde{y}(u, v, 0) \star C(u, v, w)\, ,
	\label{eq:w_projection}
\end{equation}
where $\tilde{y}(u, v, 0)$ is the Fourier transform of the sky brightness 
\begin{equation}
\begin{split}
	\tilde{y}(u, v, 0) = \int x(l, m){\rm e}^{-2\pi i \bar{w}(\sqrt{1 - l^2 - m^2} - 1)}\\
	\times{\rm e}^{-2\pi i (lu + mv)}\,  {\rm d}l{\rm d}m \, .
	\end{split}
\end{equation} 
where the projection kernel $C$ is the Fourier representation of $c$, and $\star$ is the convolution operation.
\subsection{Projection with convolutional degridding}
Since the convolution with gridding kernels is already baseline dependent, we can include the projection convolution in the gridding process.
If we let $G(u, v)$ be a gridding kernel, and the Fourier transform of the window function $g(l, m)$, we find
\begin{equation}
	\begin{split}
		y(u, v, w) =& \int \left[\frac{x(l, m)}{g(l, m)}\right]{\rm e}^{-2\pi i \bar{w}(\sqrt{1 - l^2 - m^2} - 1)} \\
		&\times g(l, m)c(l, m; w){\rm e}^{-2\pi i (lu + mv)}\,  {\rm d}l{\rm d}m \, ,
	\end{split}
\end{equation}
this suggests that we should define a new convolutional kernel
\begin{equation}
	\left[GC\right](u, v , w)= G(u, v) \star C(u, v, w)\,
\end{equation}
\begin{equation}
	y(u, v, w) = \tilde{y}(u, v, 0) \star \left[GC\right](u, v , w)\, ,
	\label{eq:w_projection_gridding}
\end{equation}
where $\tilde{y}(u, v, 0)$ is now the Fourier transform of the gridding corrected sky brightness 
\begin{equation}
\begin{split}
	\tilde{y}(u, v, 0) = \int \frac{x(l, m){\rm e}^{-2\pi i \bar{w}(\sqrt{1 - l^2 - m^2} - 1)}}{g(l, m)}\\
	\times {\rm e}^{-2\pi i (lu + mv)}\,  {\rm d}l{\rm d}m \, .
	\end{split}
\end{equation} 
Traditionally, the kernel is window separable in $l$ and $m$, i.e. $g(l, m) = g(l)g(m)$. But, as relevant for the later sections of this work, it can be a radial function, i.e. a function of $\sqrt{l^2 +m^2}$ only.			
																						
This shows that we can include the projection convolution in the gridding process through the kernel $GC$ in Equation \ref{eq:w_projection_gridding} and the operator $\bm{\mathsf{G}}\bm{\mathsf{C}}$ seen in Equation \ref{eq:matrix_equation}. In the next section, we derive expressions for the chirp kernel $C$ in $uvw$-space from a 3d setting.
																						
\section{Projection algorithm in 3d setting}
\label{sec:3dproj}
In this section, we derive the 3d $w$-projection kernel $C_{\rm H}$ formula including the horizon.
We start using a measurement equation which can be expressed to include the horizon explicitly and any restrictions of our signal to the sphere. We restrict the signal above horizon in 3d through the Heaviside step function
\begin{equation}
	\Theta(n) = 
	\begin{cases} 
		1           & n > 0 \\
		\frac{1}{2} & n = 0 \\
		0           & n < 0 
	\end{cases}
\end{equation}
and to the sphere through the Dirac delta function, yielding \mbox{$\delta(1 - l^2 - m^2 - n^2 )$},
\begin{equation}
	c_{\rm H}(l,m,n; w^\prime) = \Theta\left( n \right)\delta(1 - l^2  - m^2  - n^2 ){\rm e}^{+2\pi iw^\prime}\, .
	\label{eq:chirp_horizon}
\end{equation}
This leads to the measurement equation
\begin{equation}
	\begin{split}
		y(u, v ,w^\prime) = \int_{-\infty,-\infty,-\infty}^{\infty,\infty,\infty} x(l, m) a(l, m)c_{\rm H}(l, m, n; w^\prime)\quad\\
		\times{\rm e}^{-2\pi i (lu + mv + nw^\prime)} {\rm d}l{\rm d}m{\rm d}n\, .
	\end{split}
	\label{eq:chirp_3d}
\end{equation}
where equivalent 3d equations can be found in \citet{tho99,cor08b,tho08}. Unlike the previous section, the above equation has no $1/n$ term. This term is provided by the Dirac composition rule, which is shown in the next subsection.

\subsection{$w$-projection including the horizon directly}
In section, we show that the kernel in the work of \citet{cor08b} is equivalent to including both the horizon and spherical effects in the projection algorithm in a full 3d setting. The Fourier transform of Equation \ref{eq:chirp_horizon} is
\begin{equation}
	\begin{split}
		C_{\rm H}(u, v, w) = \int_{0,-\infty,-\infty}^{\infty,\infty,\infty} \delta(1 - l^2 -m^2 -n^2)\quad\quad\quad\quad\quad\quad \\
		\times{\rm e}^{-2\pi i (lu + mv + nw)}{\rm e}^{+2\pi iw} {\rm d}l{\rm d}m{\rm d}n\, .
	\end{split}
\end{equation}
We find that the Dirac delta function is zero at two values of $n = n_\pm$, where \mbox{$n_\pm = \pm\sqrt{1 - l^2 - m^2}$} are the two roots. In addition, we have $\delta(n^2 - n_+^2) = (\delta(n - n_+) - \delta(n - n_-))/(2n_+)$, however, the horizon eliminates the $n = n_{-}$ root from the integral. 
Using the composition rule for the Dirac delta function we have
\begin{equation}
	\begin{split}
		C_{\rm H}(u, v, w) =
		\int_{0,-1,-1}^{1,1,1} \frac{\delta(n -n_+) }{2}
		\frac{{\rm e}^{-2\pi iw\sqrt{ 1 - l^2 -m^2}}}{\sqrt{ 1 - l^2 -m^2}}\\
		\times{\rm e}^{-2\pi i (ul + mv )}{\rm e}^{+2\pi iw} {\rm d}l{\rm d}m {\rm d} n \, ,
	\end{split}
\end{equation}
where the bounds of integration are now restricted to the sphere.
and doing an integral over $n$ we find
\begin{equation}
\begin{split}
	C_{\rm H}(u, v, w) =
	\int_{-1, -1}^{1,1} \frac{{\rm e}^{-2\pi iw(\sqrt{ 1 - l^2 -m^2} - 1)}}{2\sqrt{ 1 - l^2 -m^2}}\\
	\times{\rm e}^{-2\pi i (ul + mv )} {\rm d}l{\rm d}m \, .
	\end{split}
	\label{eq:planar_chirp}
\end{equation}
This is the standard expression used for the $w$-projection kernel in \cite{cor08b}, with the inclusion of a factor of $1/2$ from there being two roots and normalization of the Dirac Delta function.
To date, there is no analytical solution for this integral beyond approximations. One reason this integral may be difficult to solve analytically, is the breaking of spherical symmetry when including the horizon. 
																						
Having no analytic solution to this integral poses a problem in understanding the properties of $C_{\rm H}(u, v, w)$. This has lead to various approximations of $C_{\rm H}(u, v, w)$, where the solution can be used estimate its support and amplitude. 
																						
We can expand $w(\sqrt{ 1 - l^2 -m^2} - 1)$ in a Taylor expansion to a given order. We can expand in \mbox{$(\sqrt{ 1 - l^2 -m^2} - 1)$} to first order, we find
\begin{equation}
	w(\sqrt{ 1 - l^2 -m^2} - 1) = -\frac{w(l^2 + m^2)}{2} + \mathcal{O}(w(l^2 + m^2)^2)\, .
\end{equation}
This has the assumption $w(l^2 + m^2)^2 \ll 1$. Also choosing a small field of view $(l^2 + m^2)^2 \ll 1$ leads to 
\begin{equation}
	\frac{{\rm e}^{-2\pi iw(\sqrt{ 1 - l^2 -m^2} - 1)}}{2\sqrt{ 1 - l^2 -m^2}} \to \frac{{\rm e}^{\pi iw(l^2 + m^2)}}{2}\, .
\end{equation}
In \cite{cor08b}, they state the above small field of view approximation, which is a Gaussian. The Fourier transform of a Gaussian function is also Gaussian, and leads to
\begin{equation}
	C_{\rm H}(u, v, w) \propto \frac{{\rm e}^{i\pi\frac{(u^2 + v^2)}{w}}}{iw}\, ,
\end{equation}
however, they comment that this expression breaks down at large fields of view and diverges at $w = 0$. 
By choosing to fix the sky to a parabola, rather than the sphere, we arrive at the same approximation above. First we choose
\begin{equation}
	c_{\rm H}(l, m, n; w^\prime) = \frac{1}{2}\delta\left(n + \frac{l^2 + m^2}{2}\right)\, ,
\end{equation}
then by integrating over $n$ in Equation \ref{eq:chirp_3d} we arrive at same small field of view approximation.
																						
\subsection{$w$-projection with exact spherical correction}
We choose to replace the horizon with a window function, where the expression for the full sphere is 
\begin{equation}
	c_{\rm H}(l, m, n; w^\prime) = h(n)\delta(1 - l^2  - m^2 - n^2)\, .
\end{equation}
Any scaling from this window function can be corrected in the upper hemisphere of the measurement equation
\begin{equation}
	\begin{split}
		y(u, v , w^\prime) = \int_{-\infty,-\infty,-\infty}^{\infty,\infty,\infty} \frac{x(l, m) a(l, m)}{h(\sqrt{1 - l^2 - m^2})} c_{\rm H}(l, m, n; w^\prime)\quad\\
		\times{\rm e}^{-2\pi i (ul + mv + nw^\prime)} {\rm e}^{+2i\pi w^\prime} {\rm d}l{\rm d}m{\rm d}n\, .
	\end{split}
\end{equation}
\subsubsection{No horizon}
When $h(n) = 1$ there is no horizon and the $w$-projection kernel is calculated from
\begin{equation}
	\begin{split}
		C(u, v, w) = \int_{-\infty,-\infty,-\infty}^{\infty,\infty,\infty} \delta(1 -l^2 -m^2 -n^2)\\
		\times{\rm e}^{-2\pi i (ul + mv + nw)} {\rm e}^{+2\pi iw} {\rm d}l{\rm d}m{\rm d}n\, .
	\end{split}
\end{equation}
The Fourier transform of this equation has an analytic solution that can be simply expressed as a real valued function
\begin{equation}
	C(u, v, w) = 2\pi{\rm sinc}(2\pi\sqrt{u^2 + v^2 + w^2}) {\rm e}^{+2\pi iw}\, ,
	\label{eq:analytic_kernel}
\end{equation}
as shown in \cite{vem61}, which is solved in spherical coordinates due to symmetry. This solution dates back as far as \citet{poi1820}, and similar problems have been solved in 2 dimensions in \citet{par1805}. The units of $(u, v, w)$ are implicitly chosen to depend on the directional cosines $(l, m, n)$, meaning $\sqrt{u^2 + v^2 +w^2} = 1$ corresponds to the largest spatial scales. 
																						
The Sinc function above represents limits on the resolution in $(u, v, w)$ due to the field of view being bounded to the sphere. The uncertainty principle states that restricting the field of view is equivalent to enforcing a resolution limit on $C(u, v, w)$. At a small field of view, this kernel is effectively a delta function of small support. However, as the field of view increases, the kernel becomes a radial Sinc function with extended support and rapid oscillations. When mosaicking multiple fields of view, resolution in $(u, v, w)$ is increased (as discussed in \cite{Ekers79} and \cite{tho99}), however, the total field of view will be limited to the sphere as represented by this radial Sinc function.
																						
Since $x(l, m)$ is independent of $n$ it will project both onto the sphere for $n$ and $-n$. While $C(u, v, w)$ models the curvature of the sphere, it allows a reflection of $x(l, m)$ for $-1\leq n< 0$. This is why a horizon window function needs to be included in the analysis.

\subsubsection{Projecting above the Horizon}
If we let $H(w)$ be the Fourier transform of $h(n)$, we find that the horizon effect can be understood through the convolution theorem
\begin{equation}
	C_{\rm H}(u, v, w) = H(w) \star C(u, v, w)\, .
\end{equation}
We can get an expression for the horizon limited $w$-projection kernel in the $(u, v, w)$ domain in terms of the $w$-projection kernel for the full sphere. Choosing $h(n) = \Theta (n)$ with \mbox{$H(w) =\frac{1}{2}\left[\delta (w) -\frac{i}{\pi w}\right]$}, we find an expression equivalent to Equation \ref{eq:planar_chirp} in the $(u, v, w)$ domain
\begin{equation}
	C_{\rm H}(u, v, w) = \frac{1}{2}C(u, v, w) - \frac{i}{2\pi}\int_{-\infty}^\infty \frac{C(u, v, t)}{w -t} {\rm d}t\, ,
	\label{eq:hilbert_transform}
\end{equation}
where the second term is a Hilbert transform of the sphere along the $w$-axis. Another equivalent expression can be found by choosing a box function $h(n) = \Pi(n + \frac{1}{2})$ for the horizon window, by setting $H(w) = {\rm e}^{i\pi w}\frac{\sin(\pi w)}{\pi w}$,
\begin{equation}
	C_{\rm H}(u, v, w) = \int_{-\infty}^\infty {\rm d}t\, {\rm e}^{i\pi t}{\rm sinc}(\pi t)C(u, v, w - t)  \, .
\end{equation}
We are not aware of an analytic solution to this convolution, which could improve understanding of the behavior of wide field effects.
																						
\subsection{Convolution with a gridding kernel}
\label{sec:uvw_convol}
\begin{figure}
	\includegraphics[width=8cm]{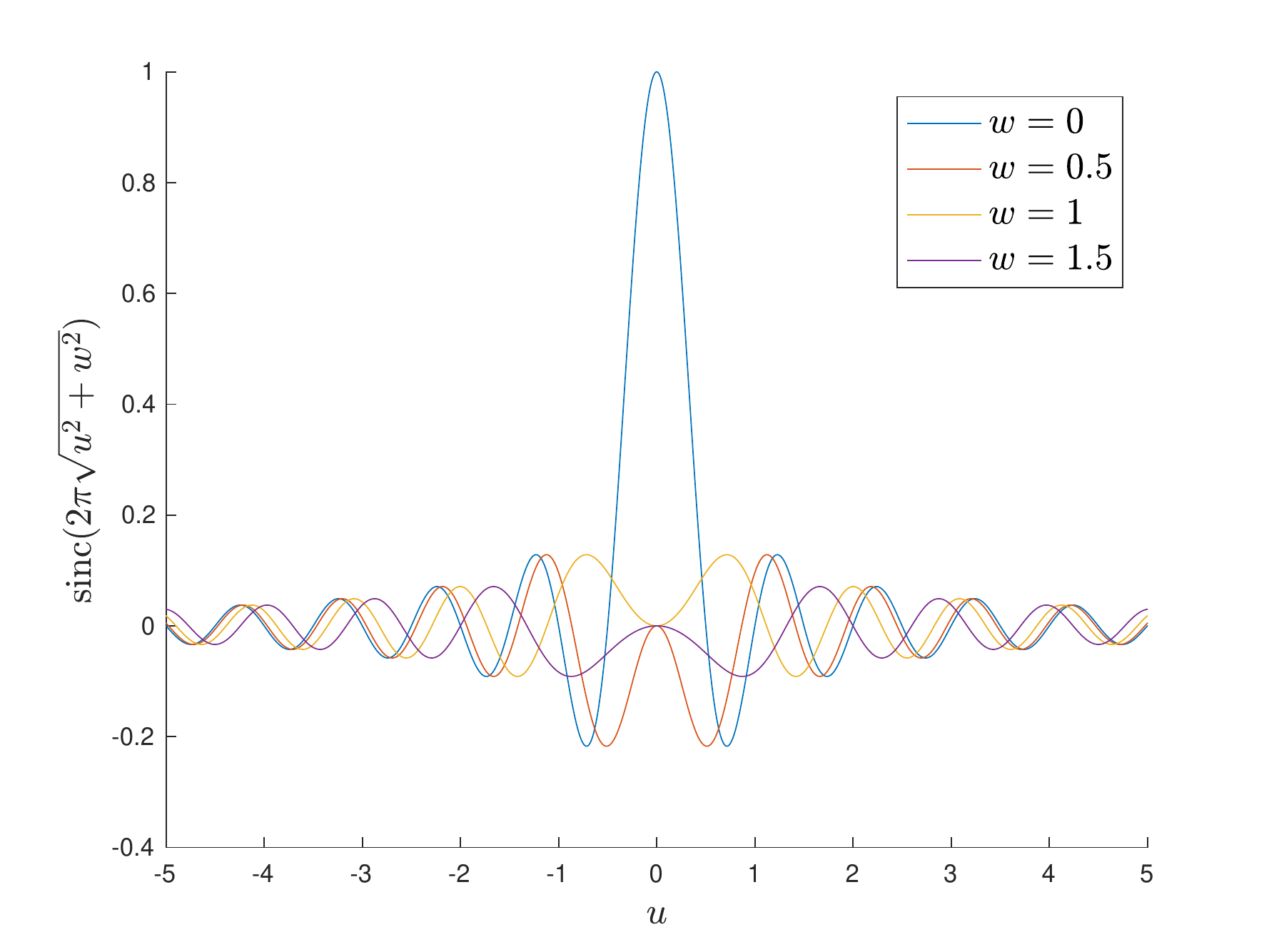}
	\includegraphics[width=8cm]{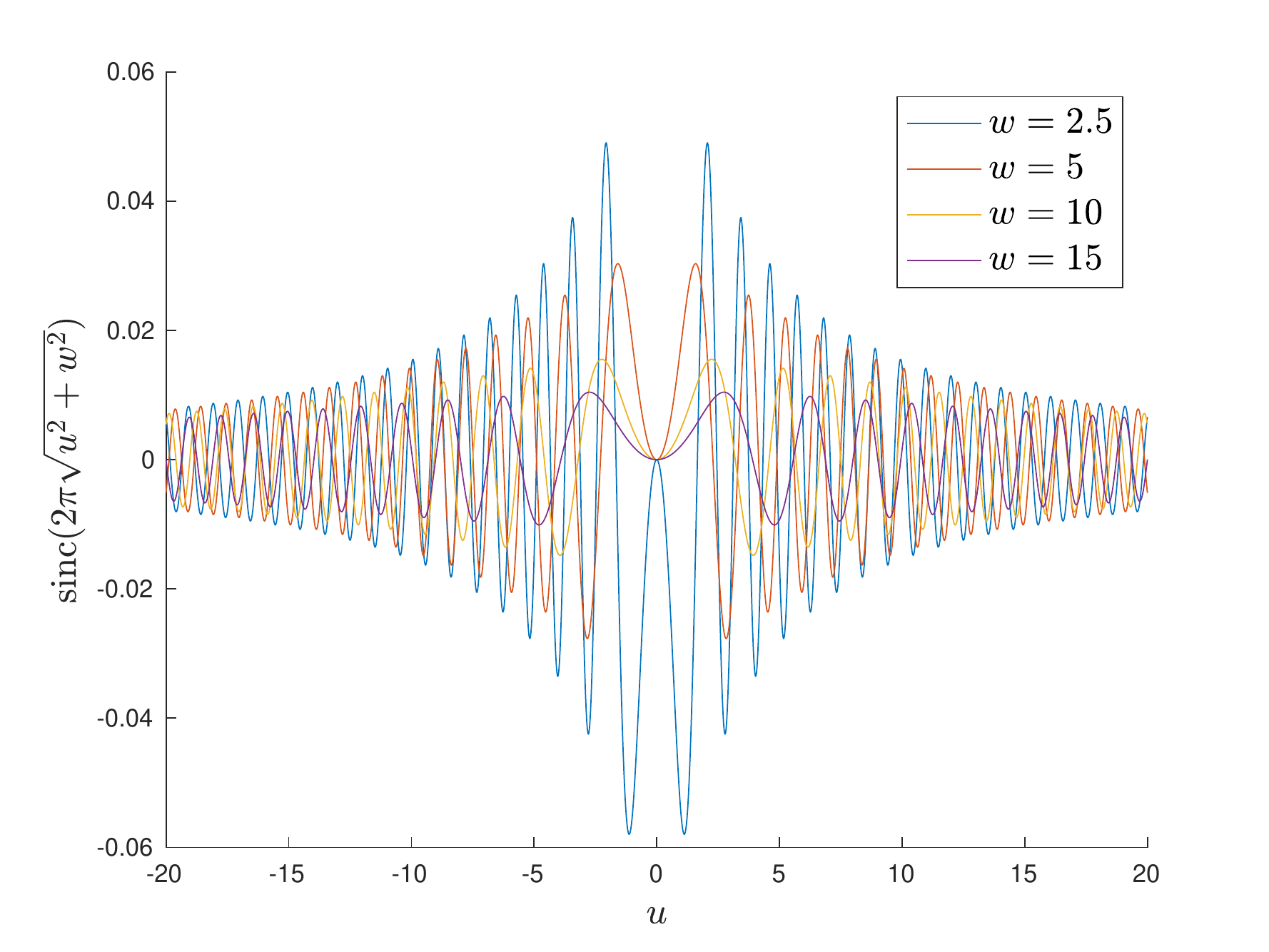}
	\caption{The oscillations of $C$, without the complex phase, as a function of $u$ for given $w$. Equation \ref{eq:uv_scale}, which is used to calculate the pixel size of a $uv$-grid, shows that many of these oscillations can occur over the convolution window, making numerical integration difficult for convolution with the gridding kernels $G$ and the horizon $H$. Hence, we find that convolution by numerical integration is difficult. Additionally, we see that $C$ has a large support that increases with $w$. The top figure shows the standard Sinc function at $w = 0$, and the bottom figure shows the spread of $C$ over a wider range of $u$ as $w$ increases.}
	\label{fig:sinc_w}
\end{figure}

To calculate the $w$-projection kernel, we could convolve the chirp with the gridding kernel in the $(u, v, w)$ domain
\begin{equation}
	\begin{split}
		[GC](u, v, w) = \int_{-\infty,-\infty,-\infty}^{\infty,\infty,\infty} G(p)G(q)H(r)\\
		\times C( u - p, v - q, w - r){\rm d}p{\rm d}q{\rm d}r\,\, .
	\end{split}
\end{equation}
However, the challenge with computing this three dimensional integral is the extended support of $H$ and $C$ in $w$. Additionally, $C(u, v, w)$ will have rapid oscillation in $(u, v)$ for small values of $w$, making accurate numerical integration and convolution expensive, see Figure \ref{fig:sinc_w}. Therefore, we avoid this approach in kernel calculation, and present an alternative approach in the next section.

\section{Kernel Calculation Methods}
\label{sec:quadrature}
In the previous section, we discussed the properties of the $w$-projection kernel in the $(l, m, n)$ and $(u, v, w)$ domains. We expected that the properties for numerical convolution with the chirp and the gridding kernel are more favorable by multiplying the window and the chirp in the image domain, then performing a Fourier transform to generate the kernel in the Fourier domain. This should increase accuracy and reduce the total computation.
																						
In this section, we describe two methods for calculating the $w$-projection kernel using the Fourier transform. The first is numerical integration using adaptive quadrature in 2d, the second is to restrict the imaged region to a radial field of view, allowing for a radially symmetric kernel that can be integrated with adaptive quadrature in 1d. In the following section we compare the numerical accuracy and speed of the two kernel construction methods. 
The scaling $\Theta(1 - l^2 - m^2)/\sqrt{ 1 - l^2 -m^2}$ is included in the gridding and primary beam correction, because it is baseline independent. We do not include this term in the gridding kernel, and we apply this in the image domain with all other baseline independent effects.

\subsection{Cartesian integration}					
To calculate the Fourier coefficients of the $w$-projection corrected gridding kernel, we need to perform a Fourier series with boundary conditions determined by the size of the window. We let $\Delta u$ and $\Delta v$ determine the conversion between pixel and baseline coordinates, $u = u_{\rm pix} \Delta u$ and $v = v_{\rm pix} \Delta v$ where $u_{\rm pix}$ and $v_{\rm pix}$ are integer pixel values. This factor is given by
\begin{equation}
	\Delta u = \left[2\alpha\sin\left(\frac{ N_x\pi{\rm cell} }{2 \times 60 \times 60 \times 180.}\right)\right]^{-1}\, .
	\label{eq:uv_scale}
\end{equation}
where cell is the size of a pixel in arc-seconds, $\alpha$ is the oversampling ratio, and $N_x$ is the image width of the $x$-axis. A similar formula is given for $\Delta v$, with respect to the $y$-axis. We use this field of view to integrate over the imaged region, and including the bounds of the sphere

\begin{widetext}
\begin{equation}
	\begin{split}
		[GC](u_{\rm pix}, v_{\rm pix}, w, \Delta u, \Delta v) =
		\int_{-\alpha/(2\Delta u), -\alpha/(2\Delta v)}^{\alpha/(2\Delta u),\alpha/(2\Delta v)}
		{\rm e}^{-2\pi iw(\sqrt{ 1 - l^2 -m^2} - 1)}
		g(\Delta u l)g(\Delta v m)\\
		\times{\rm e}^{-2\pi i (\Delta u u_{\rm pix}l + \Delta v v_{\rm pix}m )} {\rm d}l{\rm d}m\, .
	\end{split}
	\label{eq:original}
\end{equation}
We then change coordinates $l = x/\Delta u $ and $m = y/\Delta v$ to be relative to the imaged region
\begin{equation}
	\begin{split}
		[GC](u_{\rm pix}, v_{\rm pix}, w, \Delta u, \Delta v) =
		\frac{1}{\Delta u \Delta v}\int_{-\alpha/2, -\alpha/2}^{\alpha/2,\alpha/2}
		{\rm e}^{-2\pi iw(\sqrt{ 1 - x^2/\Delta u^2 -y^2/\Delta v^2} - 1)}
		g(x)g(y)\\
		\times{\rm e}^{-2\pi i (u_{\rm pix}x + v_{\rm pix}y )} {\rm d}x{\rm d}y\, .
	\end{split}
	\label{eq:analytic_convolution}
\end{equation}
\end{widetext}
Here $g(l)$ is the window function that determines the gridding kernel and $[GC]$ is the $w$-projection corrected gridding kernel. It is worth noticing that when $w = 0$, there is no dependence on $\Delta u$ or $\Delta v$, unless the condition $l^2 + m^2 \leq 1$ is to be enforced.
																											
Depending on the convention of the FFT operation $\bm{\mathsf{F}}$ in the measurement operator, there could be a phase offset of ${\rm e}^{\pm2\pi iu_{\rm pix}/2}$ and ${\rm e}^{\pm2\pi iv_{\rm pix}/2}$ required to centre the image\footnote{This is due the difference of centering the coordinates in the middle or at the corner of the image, which can require an FFT shift.}. The region of integration is determined by the zero padded field of view (we have used zero padding by a factor of $\alpha = 2$).
																											
\subsection{Polar integration}
By performing a change of coordinates, this integral can also be evaluated in polar coordinates
\begin{widetext}
\begin{equation}
	\begin{split}
		[GC](u_{\rm pix}, v_{\rm pix}, w, \Delta u, \Delta v) =
		\frac{1}{\Delta u \Delta v}\int_{0, 0}^{\alpha/2,2\pi} 
		g(r\cos(\theta))g(r\sin(\theta))
		{\rm e}^{-2\pi iw(\sqrt{ 1 - r^2\cos^2(\theta)/\Delta u^2 -r^2\sin^2(\theta)/\Delta v^2} - 1)}\\\times
		{\rm e}^{-2\pi i (u_{\rm pix}r\cos(\theta) + v_{\rm pix}r\cos(\theta) )} r{\rm d}r{\rm d}\theta\, ,
	\end{split}
	\label{eq:analytic_convolution_polar}
\end{equation}
\end{widetext}

The region is circular rather than rectangular, which is a fundamental difference with the Cartesian expression in Equation \ref{eq:analytic_convolution} (the boundary conditions for the Fourier series lie on a circle, rather than a square).

The enforces a Sinc convolution with the $w$-projection for the rectangular boundary condition, and a Airy Pattern convolution (first order Bessel Function) for the circular boundary condition. This translates to a slightly different interpolation when up-sampling the $w$-projection kernel, Sinc interpolation in the rectangular case, and $J_1(4\pi\sqrt{u^2 + v^2}/\alpha)/(2\sqrt{u^2 + v^2}/\alpha)$ interpolation in the circular case, both enforcing a band-limit.

It is important to state, this boundary is at the edge of the zero-padded region, which suggests that there would be little difference in practice because it is far outside of the gridding corrected region, and will not change suppression of aliasing error (which is the purpose of the window function/gridding convolution function). This means that while the kernels are fundamentally different due to the boundary condition, they will perform the same role, and the entire measurement operators will be equivalent after gridding correction and zero-padding.
					
\subsection{Radial symmetry}																			
We now make our window function radially symmetric \mbox{$g(l)g(m) \to g(\sqrt{l^2 + m^2})$}, and choose $\Delta u = \Delta v$ so that the chirp is also radially symmetric. This allows us to take the Fourier transform of a radially symmetric function, which is calculated using a 1d integral rather than the 2d polar integral in Equation \ref{eq:analytic_convolution_polar}, and is known as a Hankel transform\footnote{\citet{bir94} suggested that convolutions between radially symmetric functions can be efficiently computed using a Hankel Transform but in different astronomical contexts.}. This is given by
\begin{widetext}
\begin{equation}
	\begin{split}
		[GC](\sqrt{u_{\rm pix}^2 + v_{\rm pix}^2}, w, \Delta u) =
		\frac{2\pi}{\Delta u ^2}\int_{0}^{\alpha/2} g(r)
		{\rm e}^{-2\pi iw(\sqrt{ 1 - r^2/\Delta u^2} - 1)}
		J_0\left(2\pi r \sqrt{u_{\rm pix}^2 + v_{\rm pix}^2}\right) r{\rm d}r\, ,
	\end{split}
	\label{eq:analytic_convolution_hankel}
\end{equation}
\end{widetext}
where $J_0$ is a zeroth order Bessel function. The restriction of $r/\Delta u < 1$ is built into the bounds of the integration.
This has the large computational advantage of only sampling along the radius, reducing how the computation scales with field of view and $w$. There is also an increase in accuracy, since there is no sampling in $\theta$. Furthermore, the condition that we require $\Delta u = \Delta v$ is not difficult to accommodate in many cases. 
																											
\subsection{Adaptive quadrature}
To compute Equation \ref{eq:analytic_convolution}, we use adaptive multidimensional integration. In a multi-variate setting, quadrature is also known as cubature.
					
We use the software package Cubature\footnote{https://github.com/stevengj/cubature} which has implementations of these algorithms. We use the $h$-adaptive cubature method to evaluate the integrals in this work, which uses the work of \citet{gen80} and \citet{ber91} to perform integration using an adaptive mesh to approximate the integral, until convergence is reached ($h$ is in reference to a length parameter of the mesh). Cubature also has a $p$-adaptive method \citep{ern89}, which uses polynomial based quadrature, increasing the polynomial order of the integrand until the integration has converged, and is expected to converge faster than $h$-adaptive methods for smooth integrands. 

The $p$-adaptive would converge faster than the $h$-adaptive method for the 1d-integration, while providing results as accurate within numerical error. However, the accuracy of the $p$-adaptive method was not as accurate for 2d-integration, especially in the presence of discontinuities. For this reason, we use the $p$-adaptive method for 1d-integration but the $h$-adaptive method for 2d-integration.

\subsection{Kaiser-Bessel gridding kernel} 
In this work, we use a Kaiser-Bessel gridding kernel. Kaiser-Bessel functions have been used as convolutional gridding kernels for decades \citep{gre79,jac91,fes03}, and have a simpler form than the prolate spheroidal wave functions, while providing similar performance \citep{gre79}. 
The zeroth order Kaiser-Bessel function can be expressed as
\begin{equation}
	G(u_{\rm pix}) = \frac{I_0\left(\beta \sqrt{1 -\left(\frac{2 u_{\rm pix}}{J}\right)^2}\right)}{I_0(\beta)}\, ,
\end{equation}
where $u_{\rm pix}$ has units of pixels, $J$ is the support in units of pixels, $I_0$ is the zeroth order modified Bessel function of the first kind, and $\beta$ determines the spread of the Kaiser-Bessel function \citep{jac91,fes03}.
The Fourier Transform of $G(u_{\rm pix})$ is
\begin{equation}
	g(x)= {\rm sinc}\left(\sqrt{\pi^2 x^2 J^2 - \beta^2}\right)\, .
\end{equation}
To correct for the convolution, the image is divided by $g(l)$ \citep{jac91,fes03}
\begin{equation}
	s(x) = \left[g(x)\right]^{-1}\, .
\end{equation}
The work of \citet{fes03} shows that for $\beta = 2.34 J$ the Kaiser-Bessel kernel performs similarly to the optimal min-max kernel considered. 
																											
In this work, we use the Kaiser-Bessel gridding kernel to calculate $w$-projection kernels, by using $g(x)$ in Equations \ref{eq:analytic_convolution} and \ref{eq:analytic_convolution_hankel}. For other possible window functions and anti-aliasing kernels, see \citet{tho08} and \citet{LP18}.
																											
\section{Validation of Radially Symmetric Kernel}
\label{sec:numerical_tests}
In this section we numerically evaluate Equation \ref{eq:analytic_convolution}, and present a cross section of the kernel, showing its variation with sub-pixel accuracy. We then numerically evaluate Equation \ref{eq:analytic_convolution_hankel}, showing that it provides the same accurate sub-pixel accuracy, with orders of magnitude less function evaluations during the quadrature computation.
					
\subsection{Quadrature convergence conditions}															
The kernel function is normalized to one when $(u, v, w) = (0, 0, 0)$, and an estimate error tolerance $\eta$ on the quadrature calculated kernel $[GC]^\eta(u_{\rm pix}, v_{\rm pix}, w)$ is used for quadrature convergence of the kernel, such that the absolute difference is less than $\eta$ 
\begin{equation}
	|[GC](u_{\rm pix}, v_{\rm pix}, w)- [GC]^\eta(u_{\rm pix}, v_{\rm pix}, w)| \leq \eta\, .
\end{equation}
It is also possible to use the relative difference
\begin{equation}
	\frac{|[GC](u_{\rm pix}, v_{\rm pix}, w)- [GC]^\eta(u_{\rm pix}, v_{\rm pix}, w)|}{|[GC]^\eta(u_{\rm pix}, v_{\rm pix}, w)|} \leq \eta \, ,
\end{equation}
which would constrain smaller values of $[GC]^\eta(u_{\rm pix}, v_{\rm pix}, w)$ to be calculated more accurately, at the cost of more computation. 
																											
There is a downside of using absolute difference, for example, if you are calculating kernels to an absolute accuracy of $10^{-2}$ and the kernels have values below $10^{-2}$ then these values may not be accurate. The relative difference is an ideal alternative, but it can cause an inconsistent level of accuracy across the measurement operator, and more computation can go into small values that may not contribute much in practice. If the support size is known accurately before computation, this may help. 

We assume that the support size of the $w$-projection $GC$ kernel is proportional to $2w/\Delta u$ and at least the support size of the gridding kernel $G$. With the support size known, we use the absolute different criteria with $\eta = 10^{-6}$.

\subsection{Kernel cross-section}																			
\begin{figure*}
	\begin{minipage}{1.0\textwidth}
		\includegraphics[width=6.2cm]{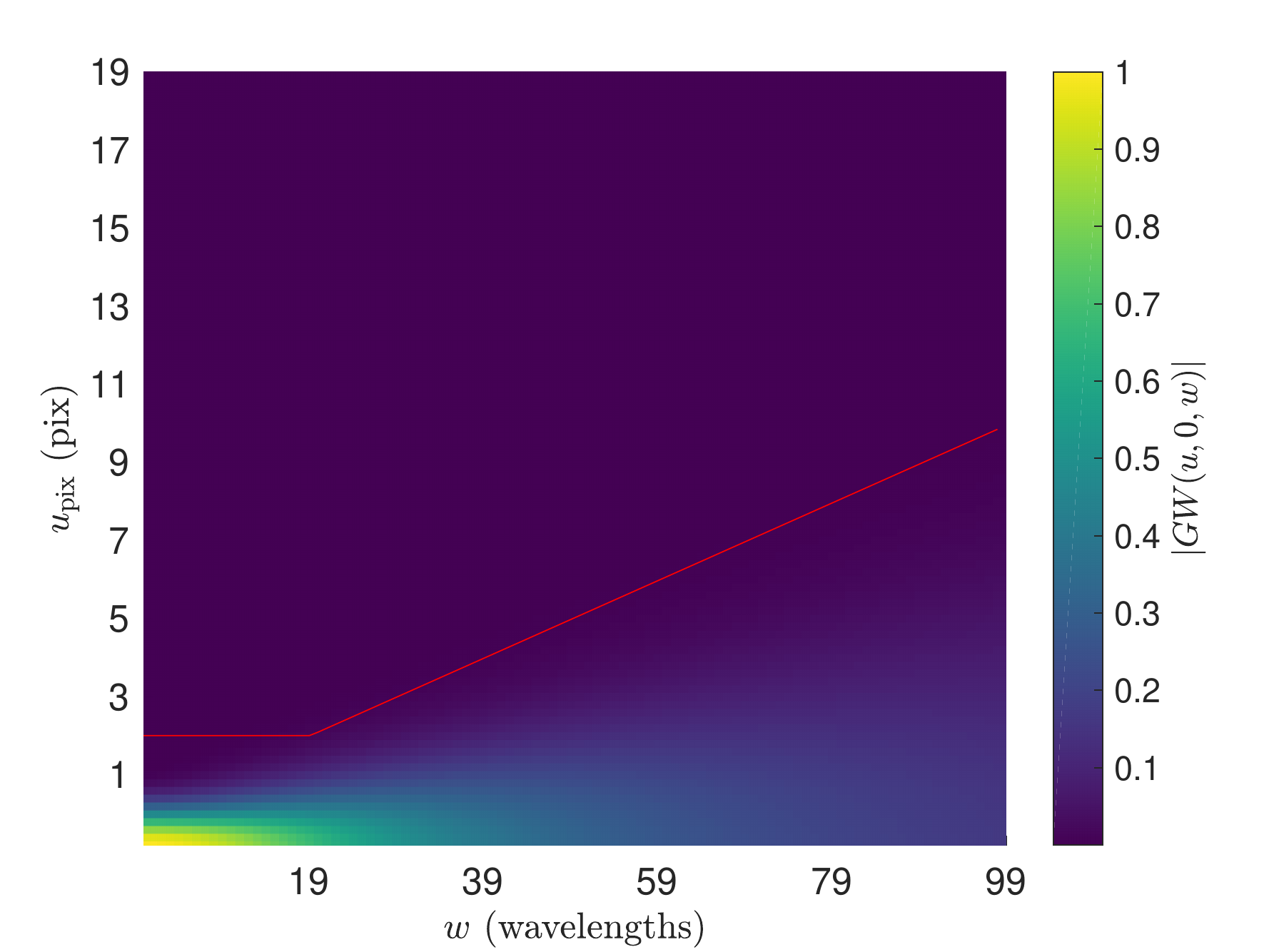}
		\includegraphics[width=6.2cm]{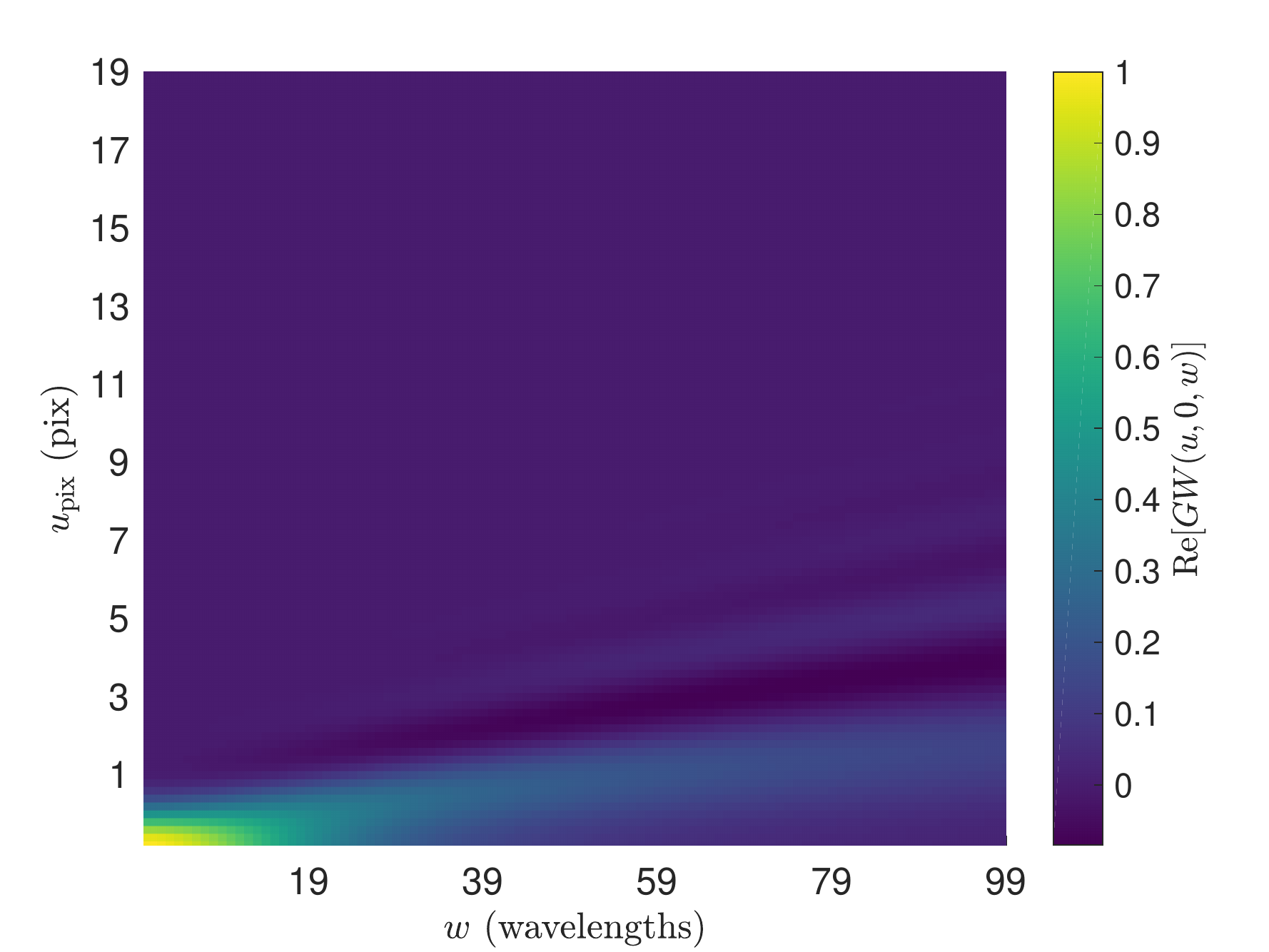}
		\includegraphics[width=6.2cm]{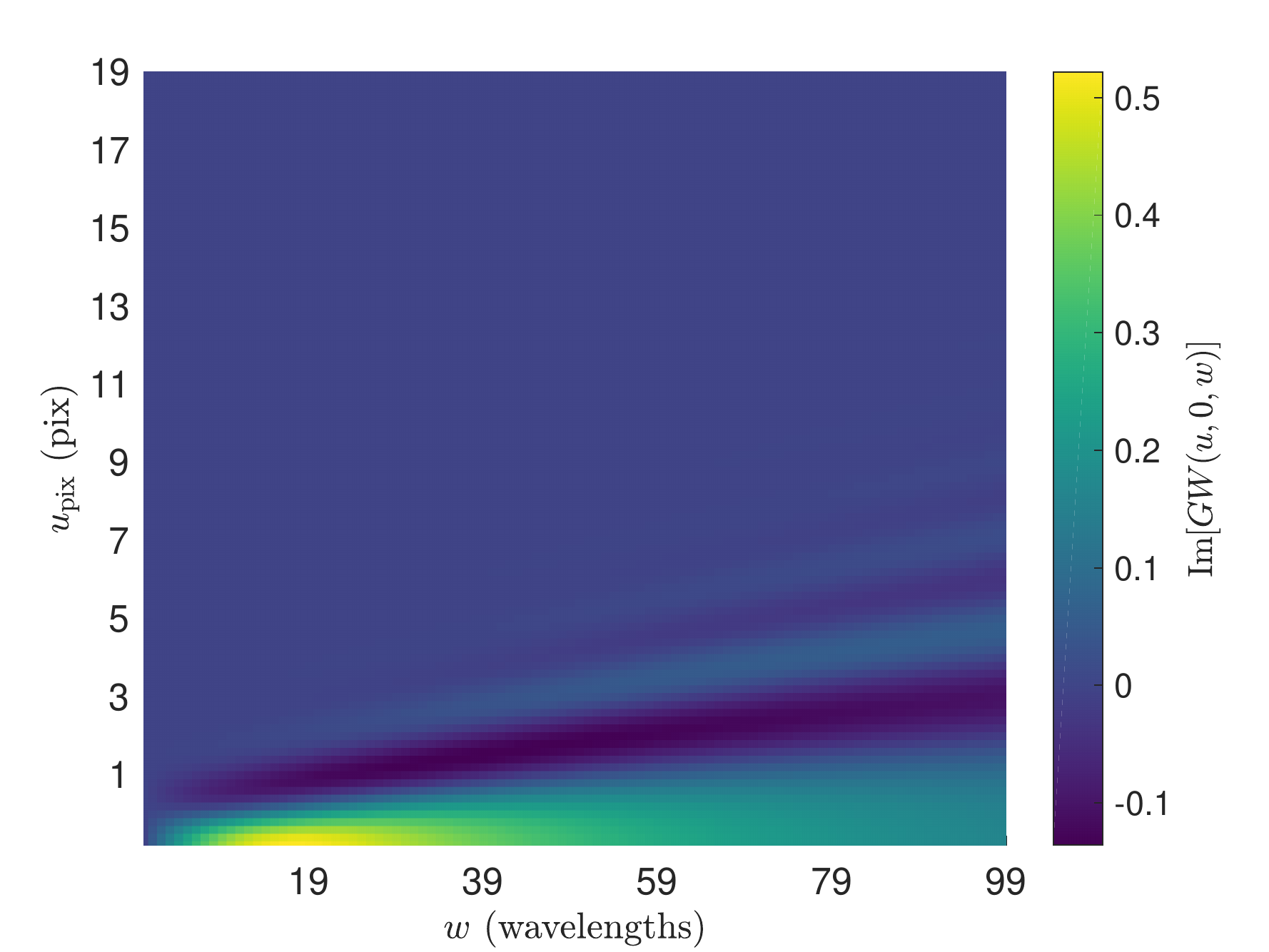}
		\includegraphics[width=6.2cm]{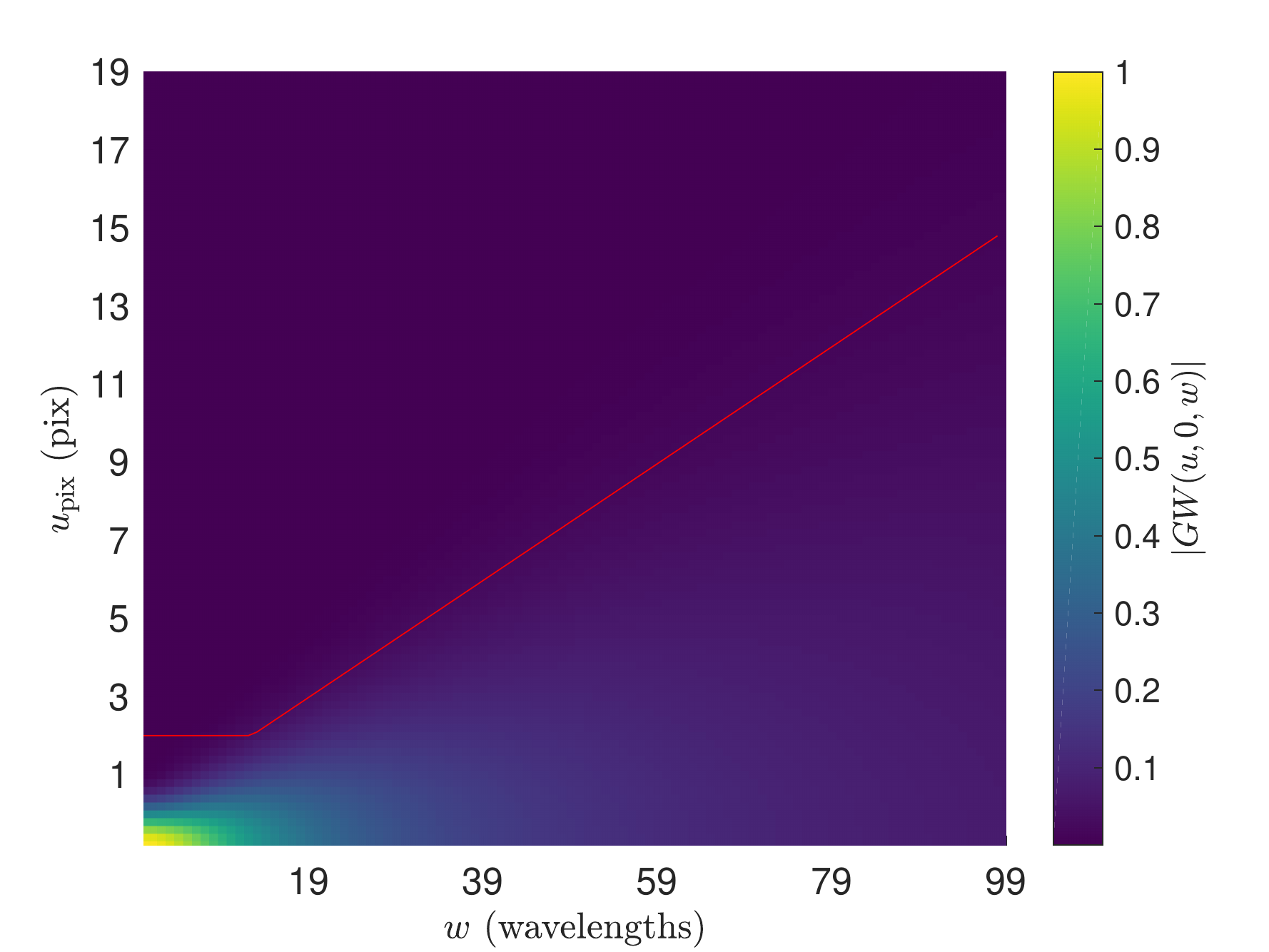}
		\includegraphics[width=6.2cm]{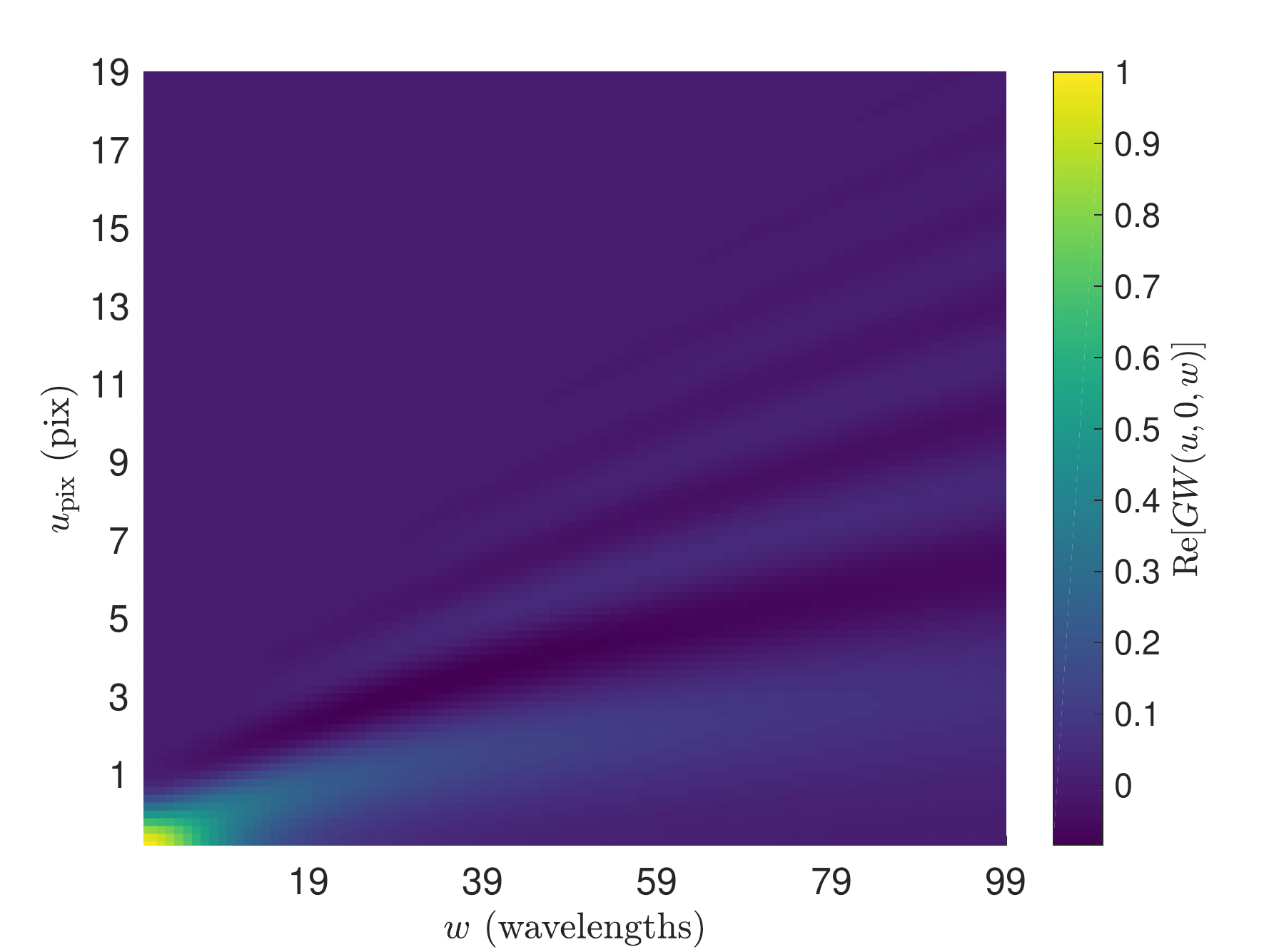}
		\includegraphics[width=6.2cm]{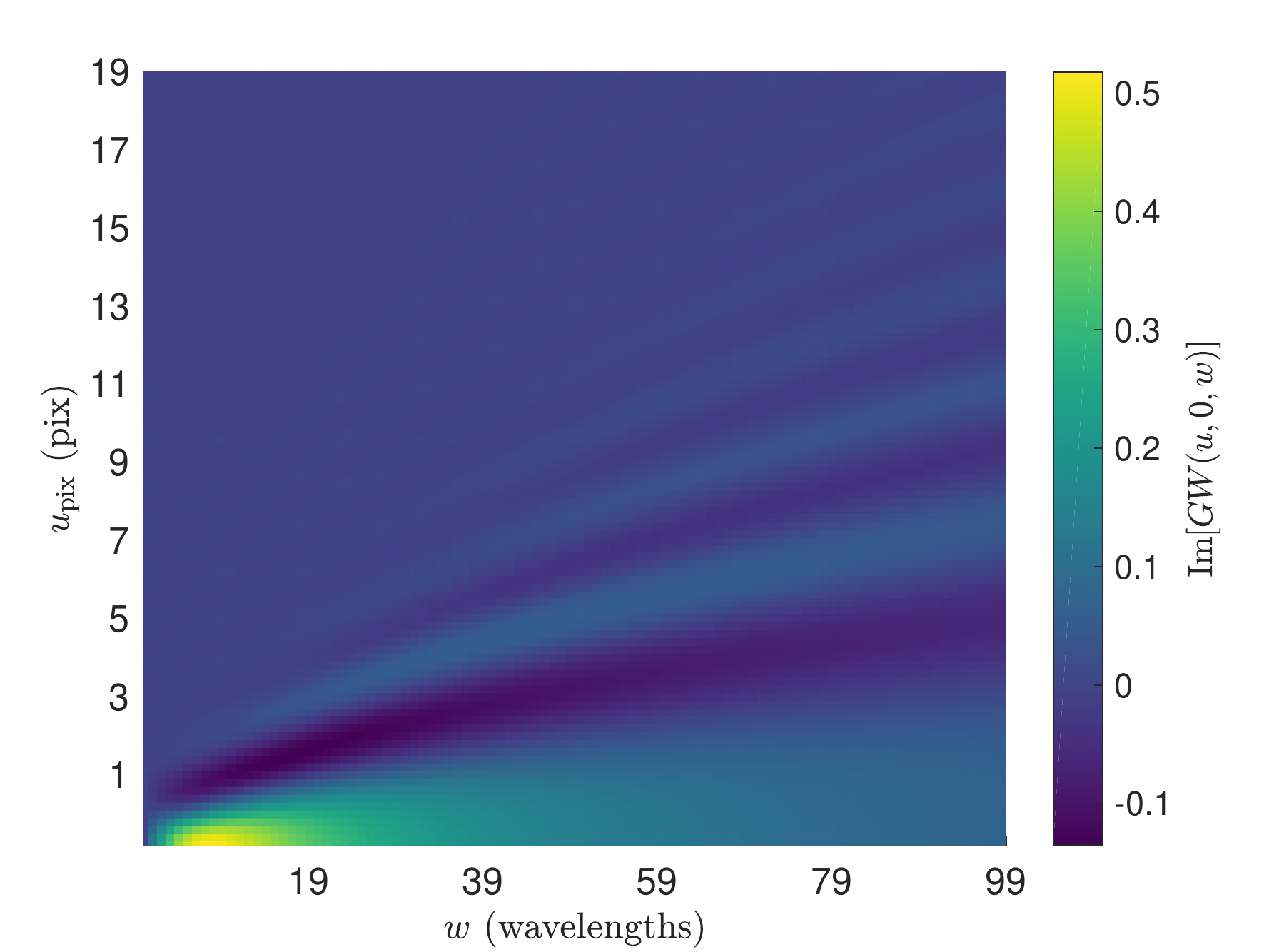}
		\includegraphics[width=6.2cm]{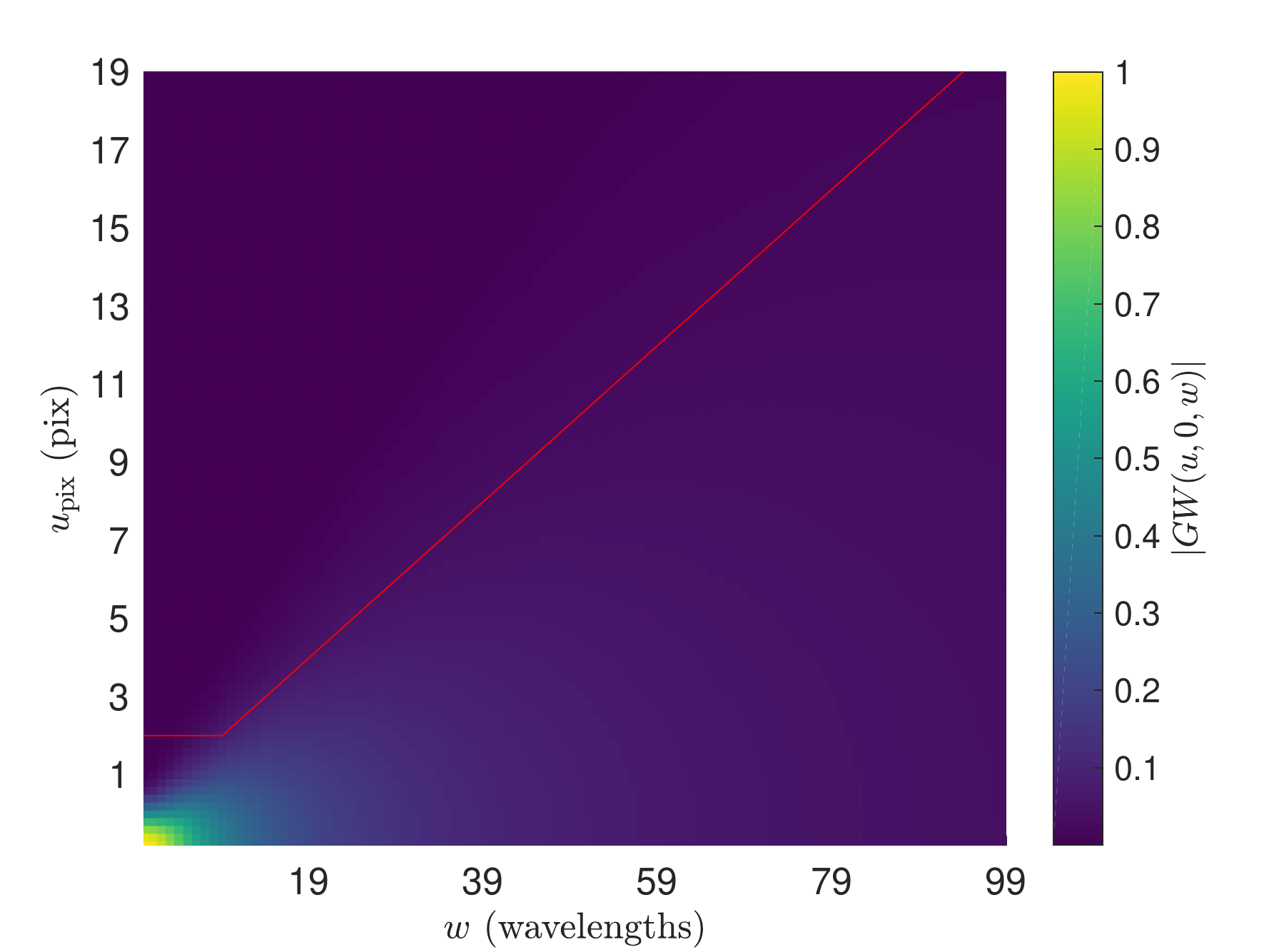}
		\includegraphics[width=6.2cm]{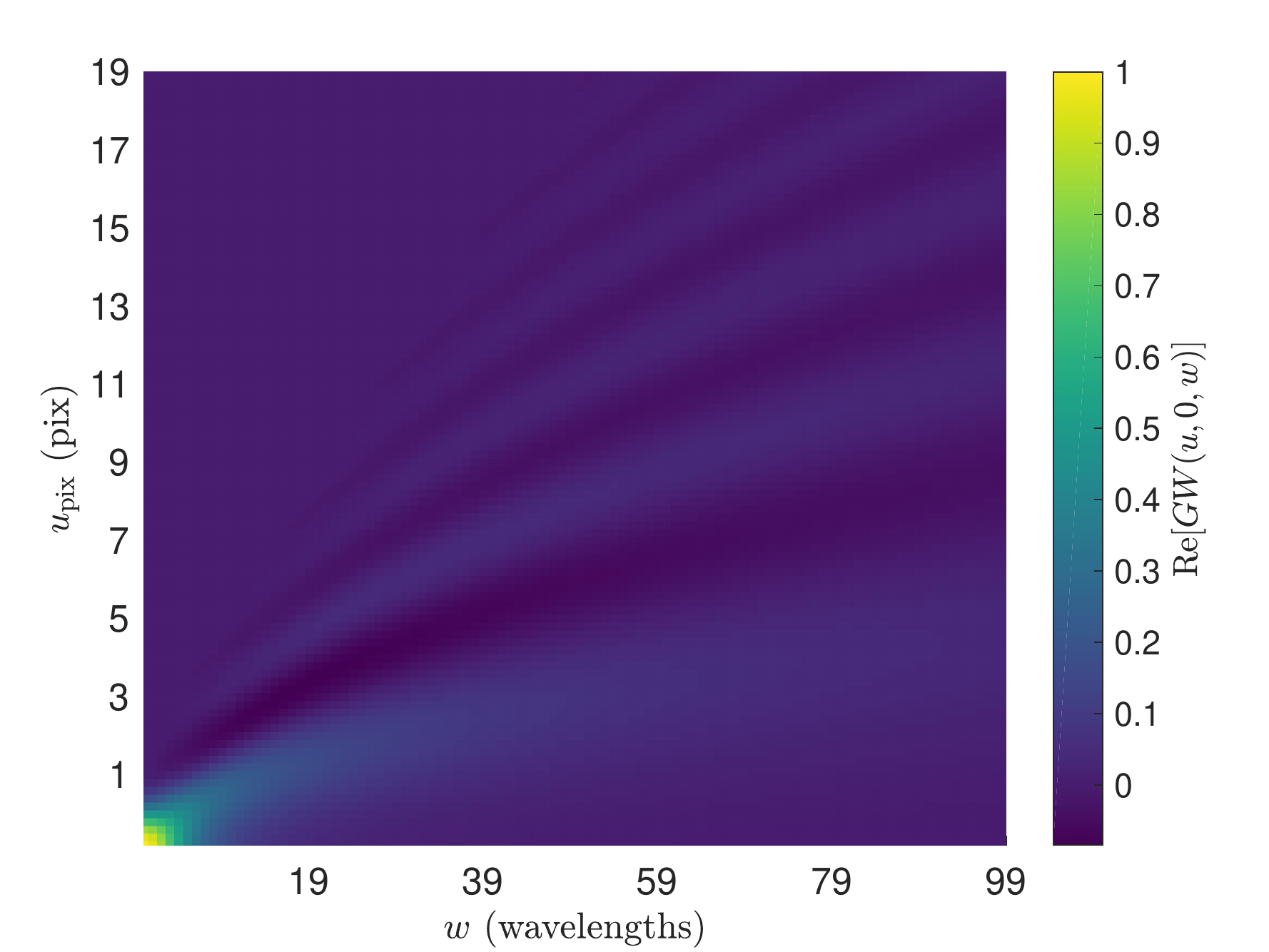}
		\includegraphics[width=6.2cm]{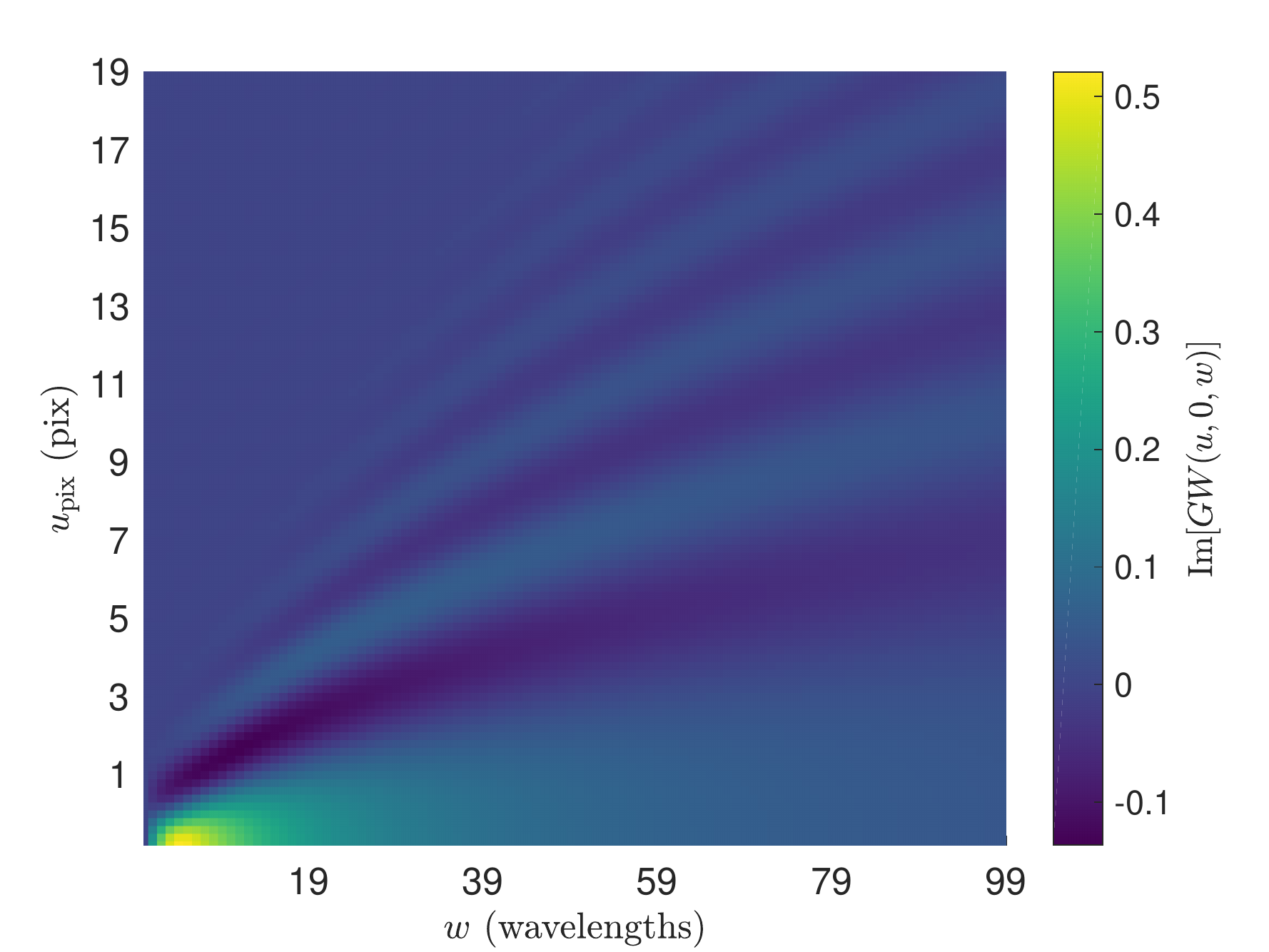}
		\caption{Plot of the kernels calculated using Equation \ref{eq:analytic_convolution}, as a function of $u_{\rm pix}$ and $w$, with $v_{\rm pix} = 0$, for absolute (left column), real (middle column), and imaginary (right column) values. Each row has a different field of view, {$11.3778^\circ \times 11.3778^\circ$} (top), {$17.0667^\circ\times 17.0667^\circ$} (middle), and {$22.7556^\circ\times22.7556^\circ$} (bottom). We see that the kernel spreads as a function of increasing $w$. The support size in pixels increases with field of view, due to a large field increasing the sampling rate of the kernel. It is also clear that the kernel decreases in value with increasing $w$, faster at wider fields of view. The real and imaginary components both show oscillations. We find the imaginary component is zero at $w = 0$ as expected. The values have been calculated using adaptive quadrature within an absolute error of $\eta = 10^{-6}$. There are 100 uniform samples in each of $u_{\rm pix}$ and $w$, making $10^4$ for each plot. The red line shows $\max (4 , 2 w/\Delta u)/2$ for reference, which is assumed to be the support size for this work. The features of this kernel are also consistent with $w$-projection kernels used by ASKAPSoft \citep{corn11}.}
		\label{fig:kernels}
	\end{minipage}
\end{figure*}

Figure \ref{fig:kernels} shows a cross section of the $w$-projection kernel $[GC](u_{\rm pix}, 0, w)$, the real and imaginary components, and the absolute value, for $ 0 \leq u_{\rm pix} \leq 19$ and $0 \leq w \leq 99$. We find that the convolution of $C_{\rm H}$ with $G(u)$ and $G(v)$ creates a smooth varying $w$-projection kernel in both real and imaginary components. The imaginary component is zero at $w = 0$, which is consistent with Equation \ref{eq:hilbert_transform}. We find that the decay in the kernel as a function of $w$ is more extreme with wider fields of view.	

\begin{figure*}
	\begin{minipage}{1.0\textwidth}
		\includegraphics[width=6.2cm]{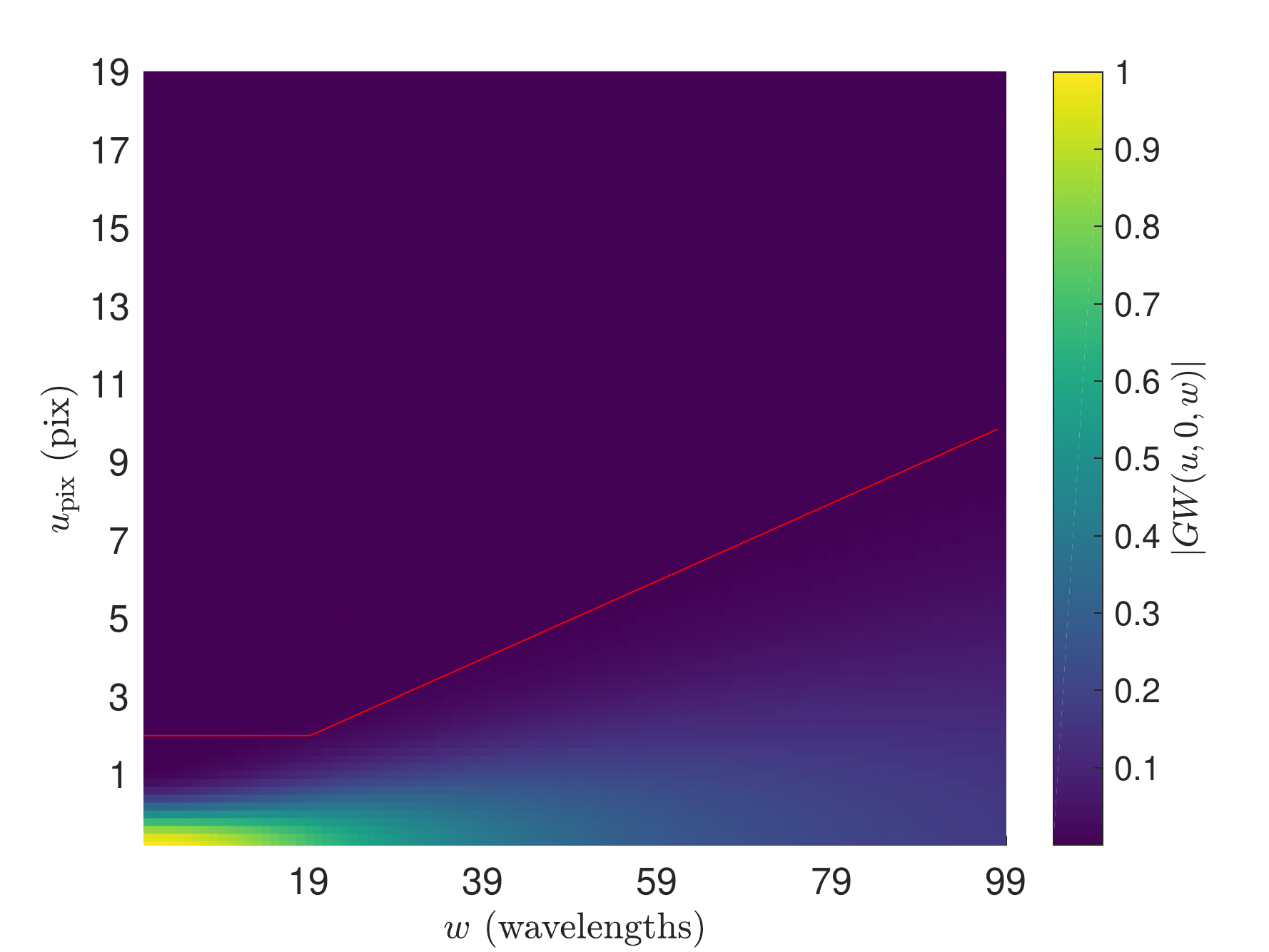}
		\includegraphics[width=6.2cm]{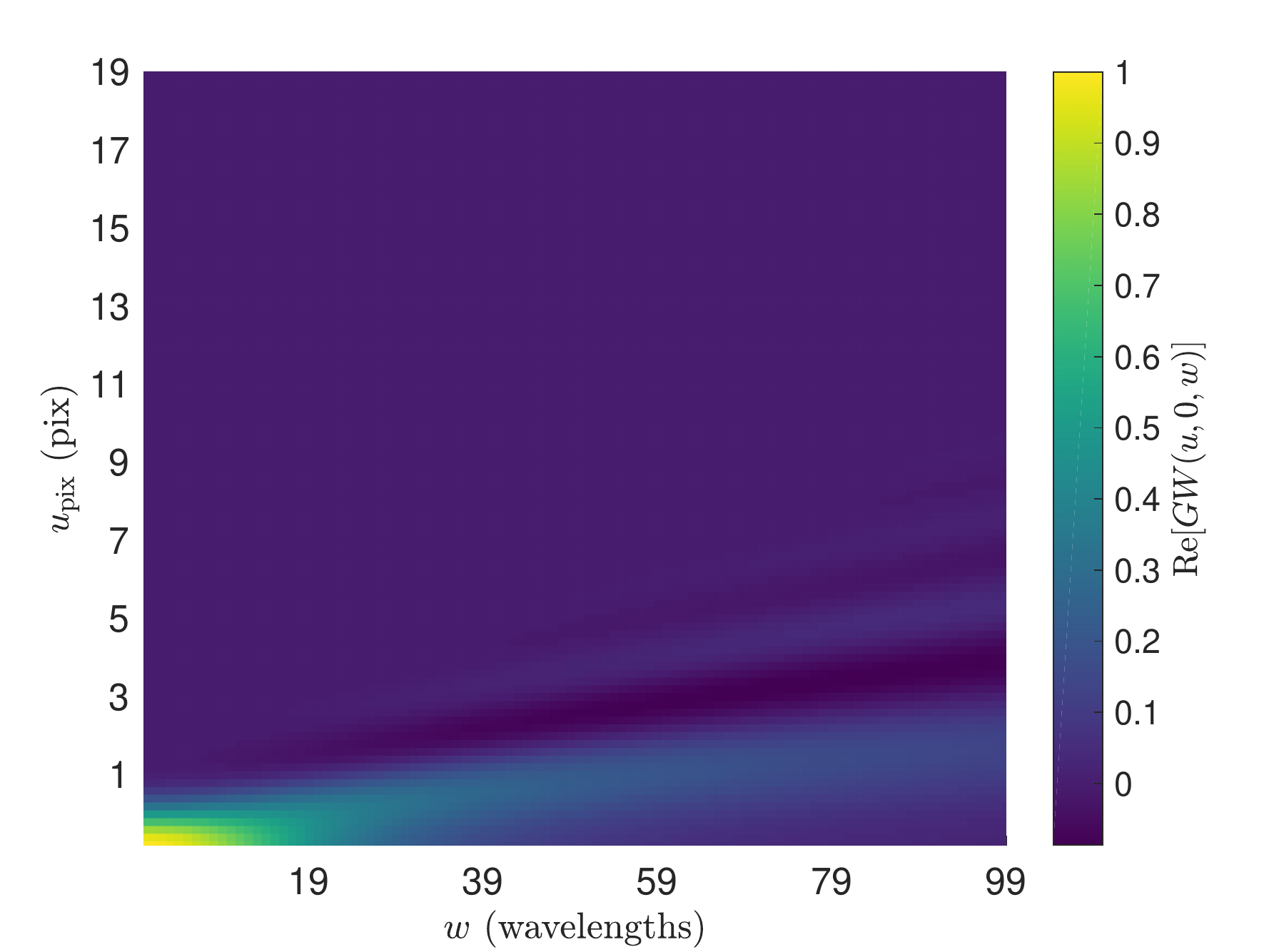}
		\includegraphics[width=6.2cm]{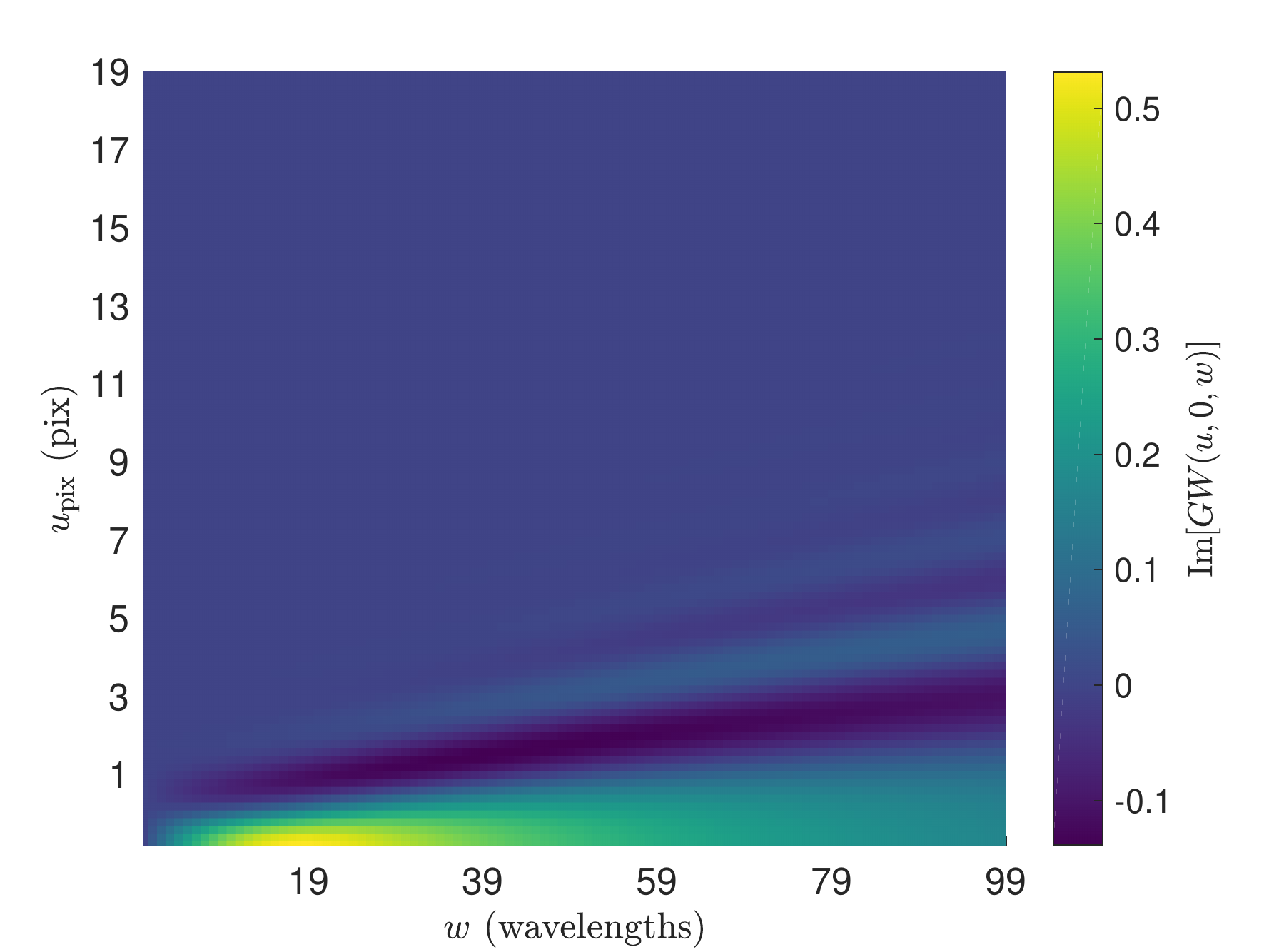}
		\includegraphics[width=6.2cm]{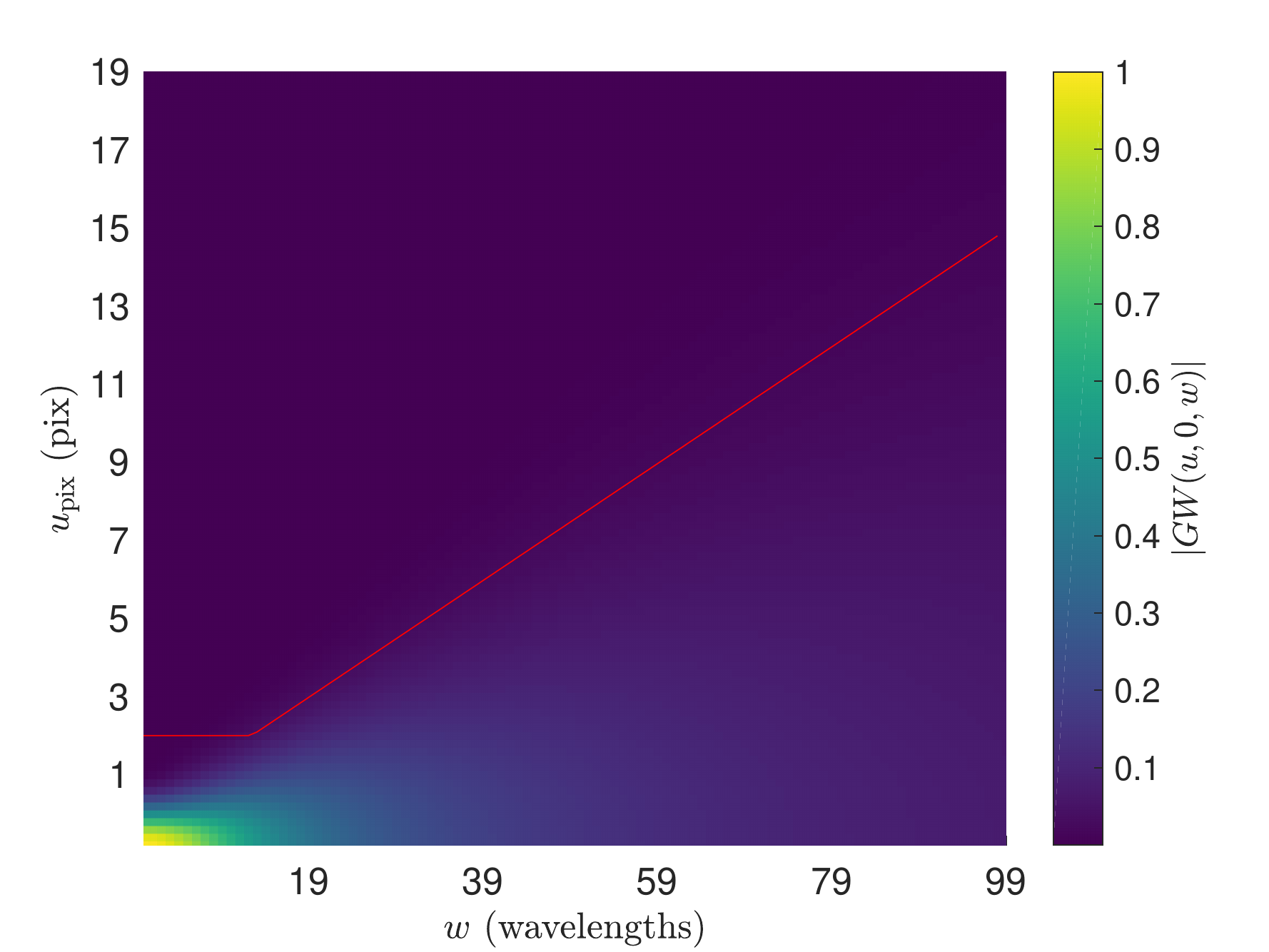}
		\includegraphics[width=6.2cm]{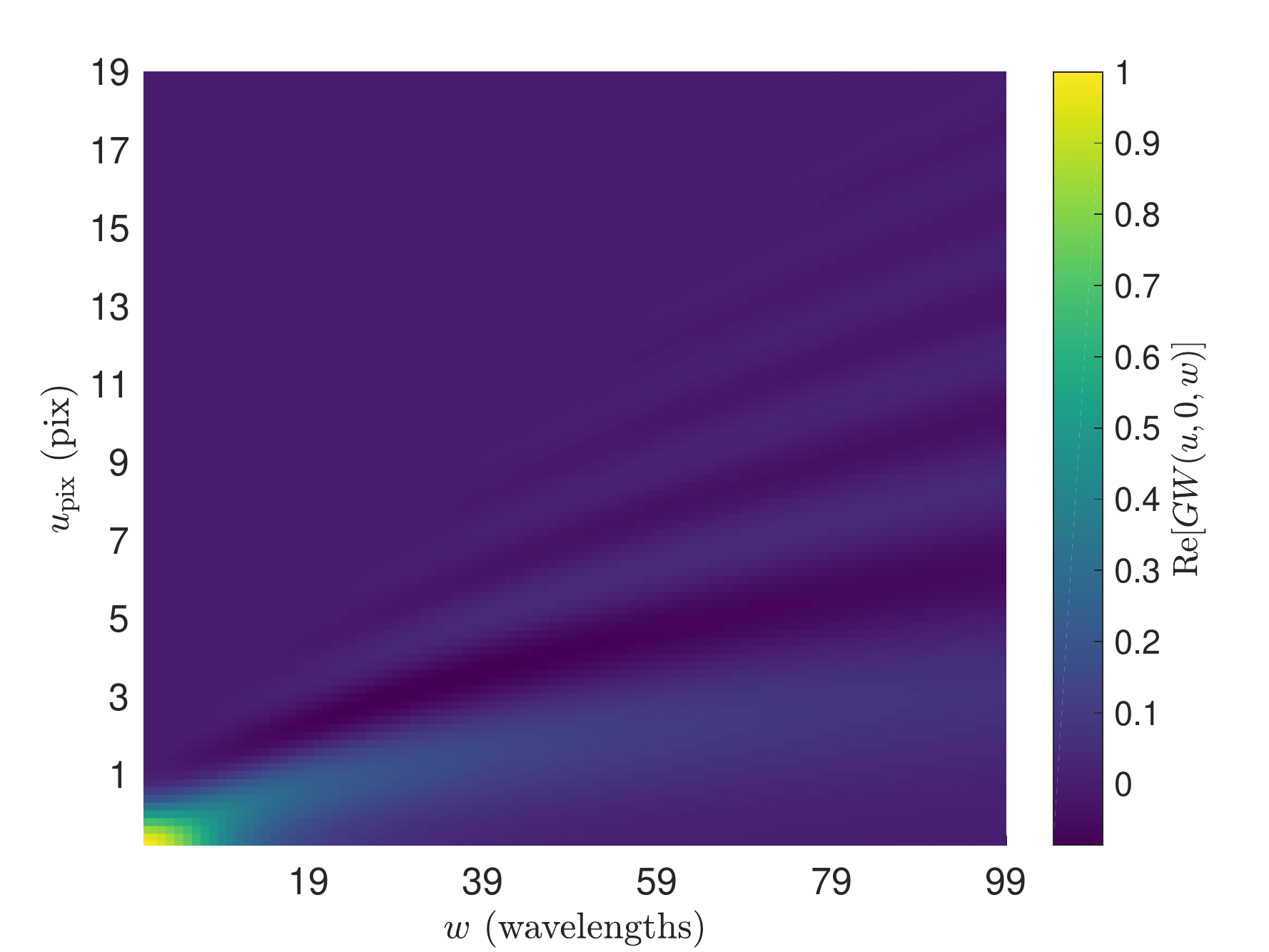}
		\includegraphics[width=6.2cm]{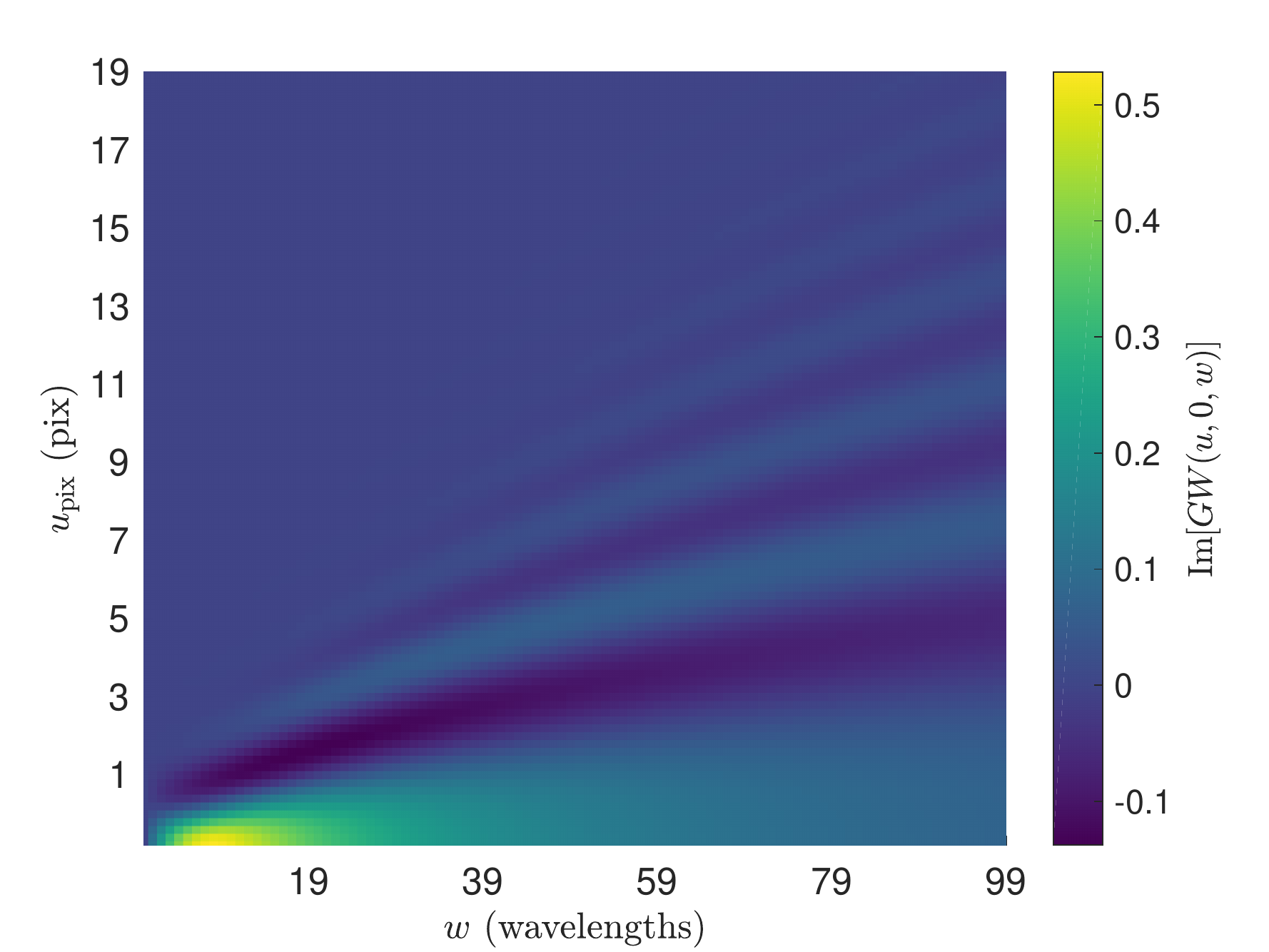}
		\includegraphics[width=6.2cm]{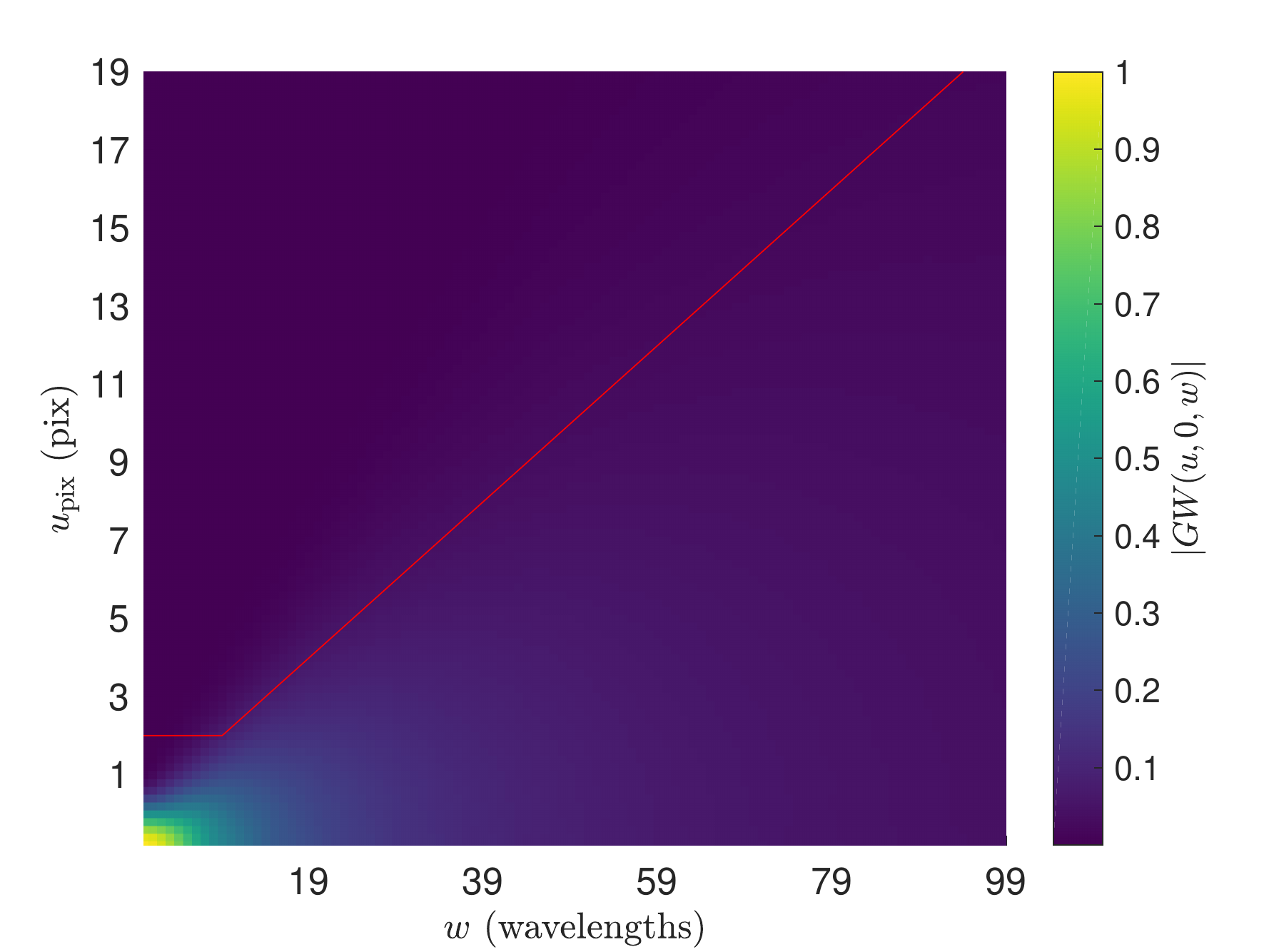}
		\includegraphics[width=6.2cm]{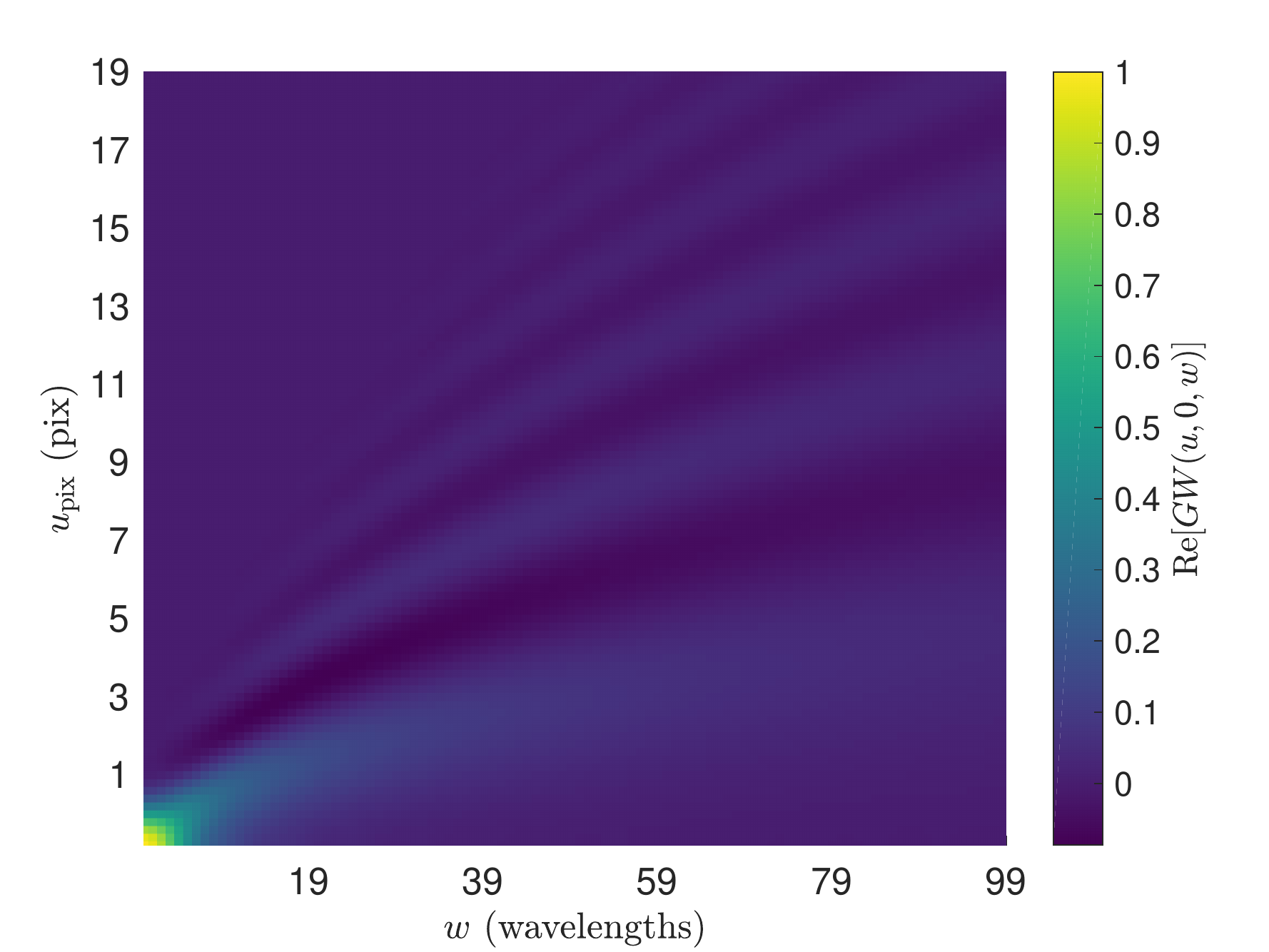}
		\includegraphics[width=6.2cm]{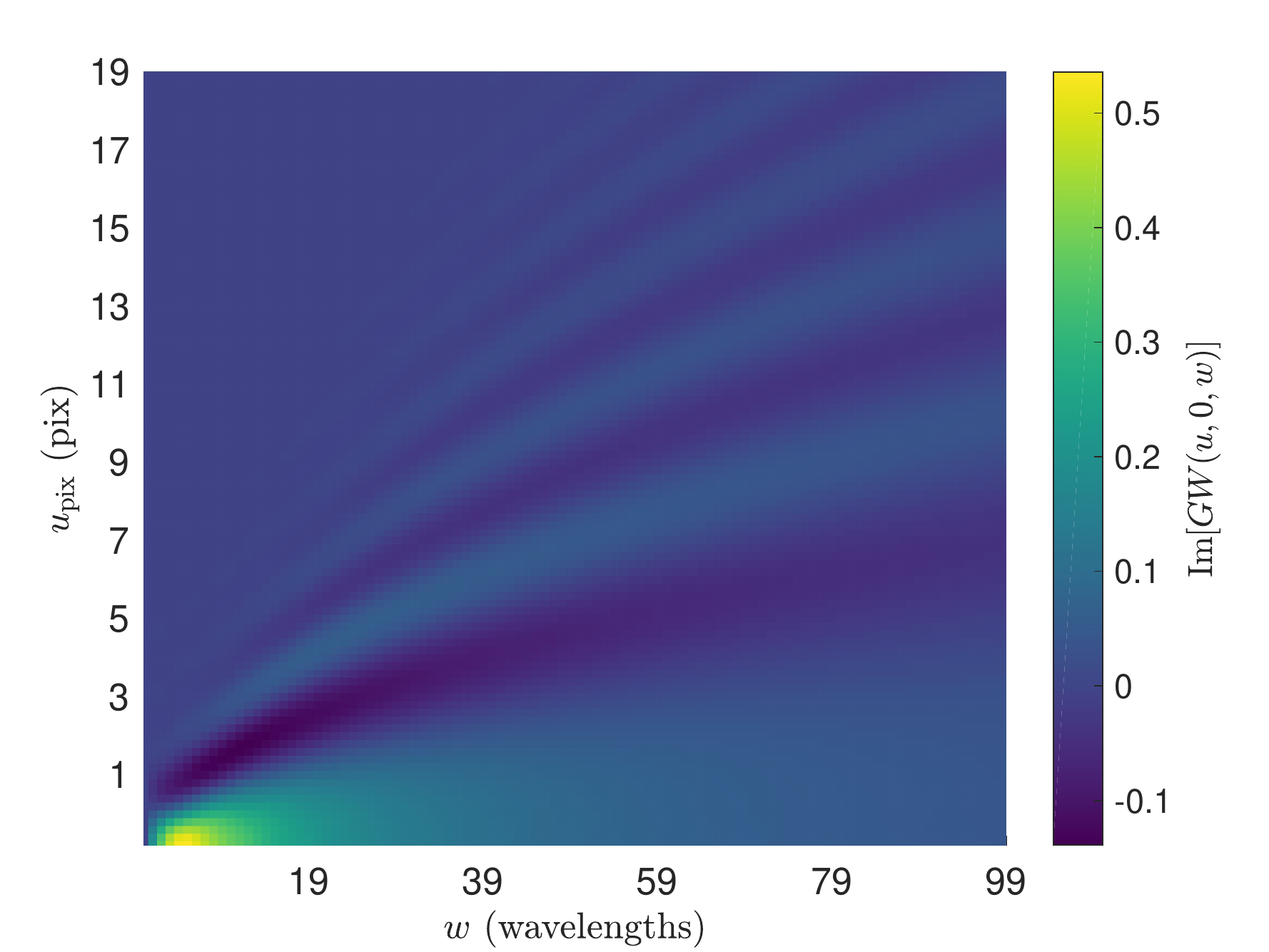}
		\caption{Plot of the kernels calculated from Equation \ref{eq:analytic_convolution_hankel}, as a function of $u_{\rm pix}$ and $w$, with $v_{\rm pix} = 0$, for absolute (left column), real (middle column), and imaginary (right column) values. Each row has a different field of view, {$11.3778^\circ\times11.3778^\circ$} (top), {$17.0667^\circ\times17.0667^\circ$} (middle), and {$22.7556^\circ\times22.7556^\circ$} (bottom). We find the same features in Figure \ref{fig:kernels}, showing that it is consistent with Equation \ref{eq:analytic_convolution}. The values have been calculated using adaptive quadrature within an absolute error of $\eta = 10^{-6}$. There are 100 uniform samples in each of $u_{\rm pix}$ and $w$, making $10^4$ for each plot. The red line shows $\max (4 , 2 w/\Delta u)/2$ for reference.}
		\label{fig:kernels_hankel}
	\end{minipage}
\end{figure*}

\begin{figure*}
	\begin{minipage}{1.0\textwidth}
		\includegraphics[width=6.2cm]{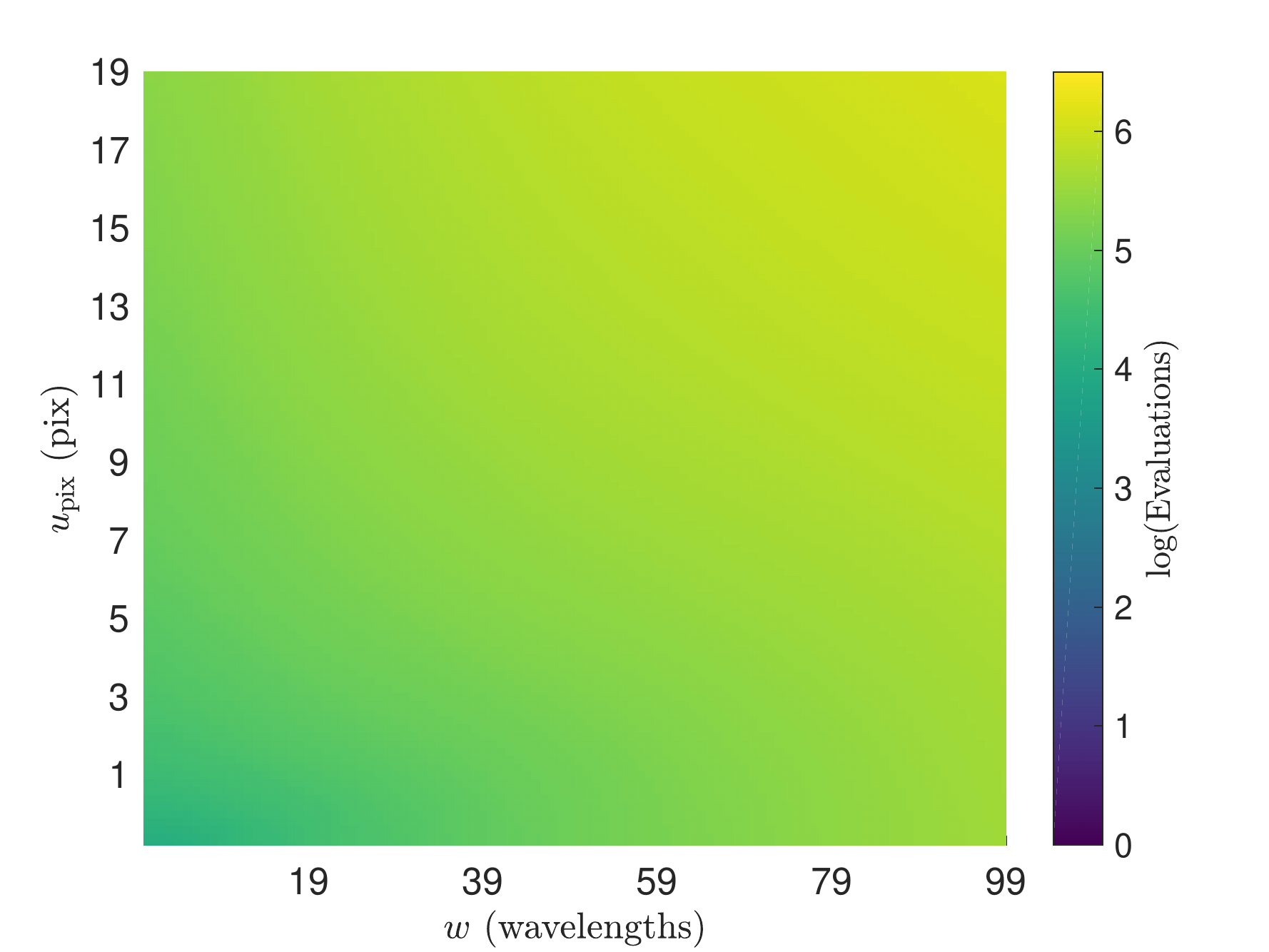}
		\includegraphics[width=6.2cm]{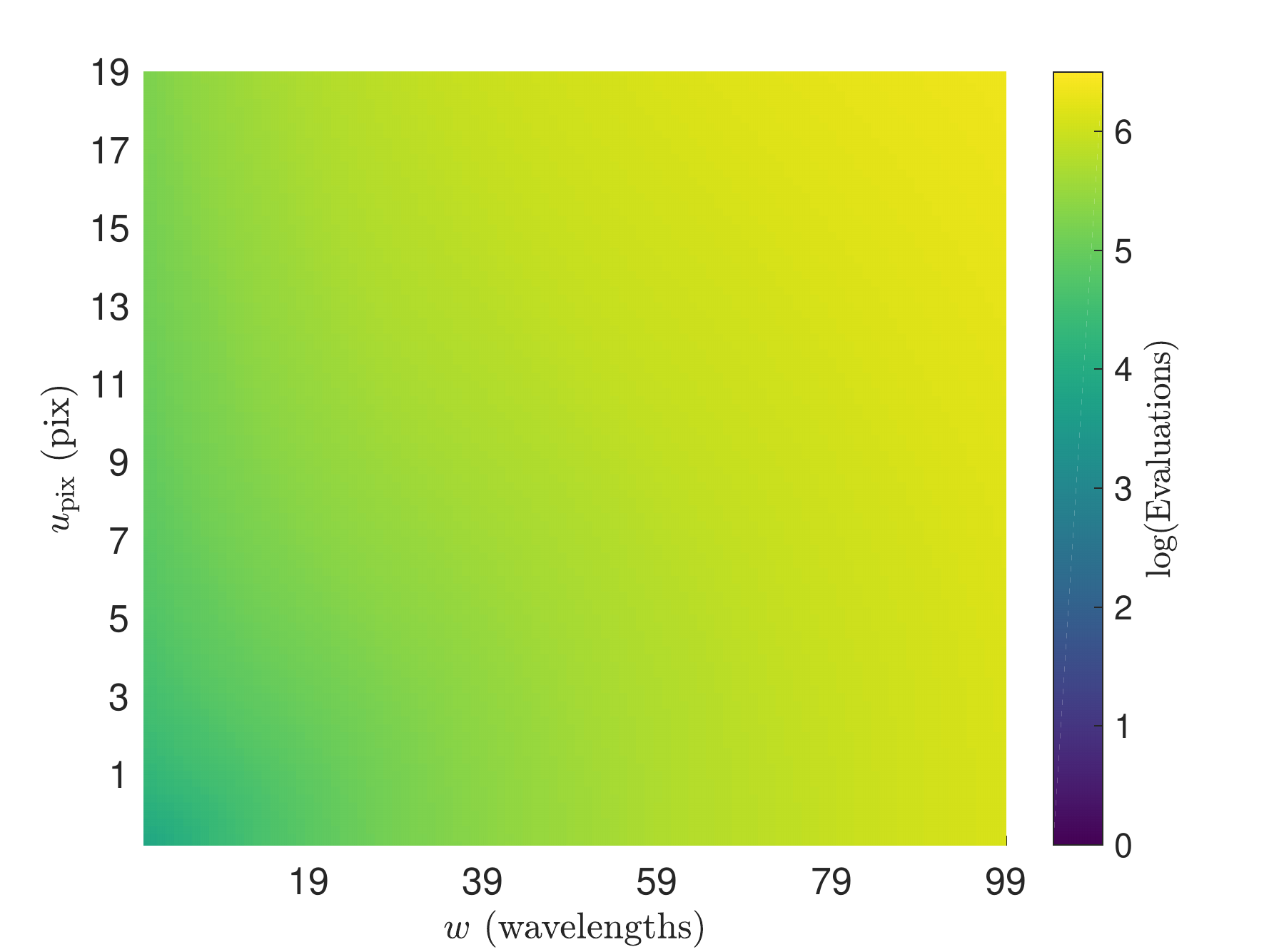} 
		\includegraphics[width=6.2cm]{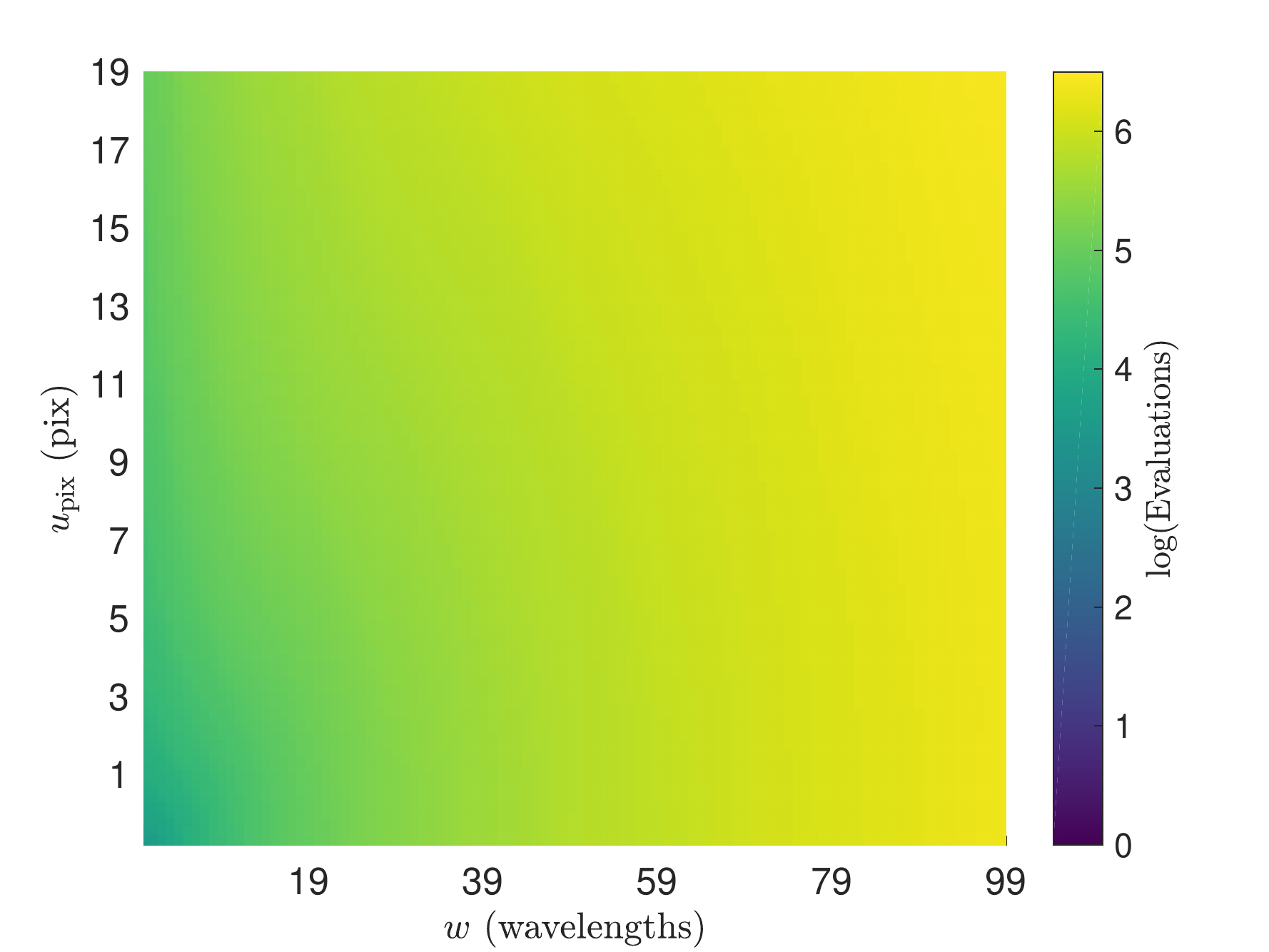}
		\includegraphics[width=6.2cm]{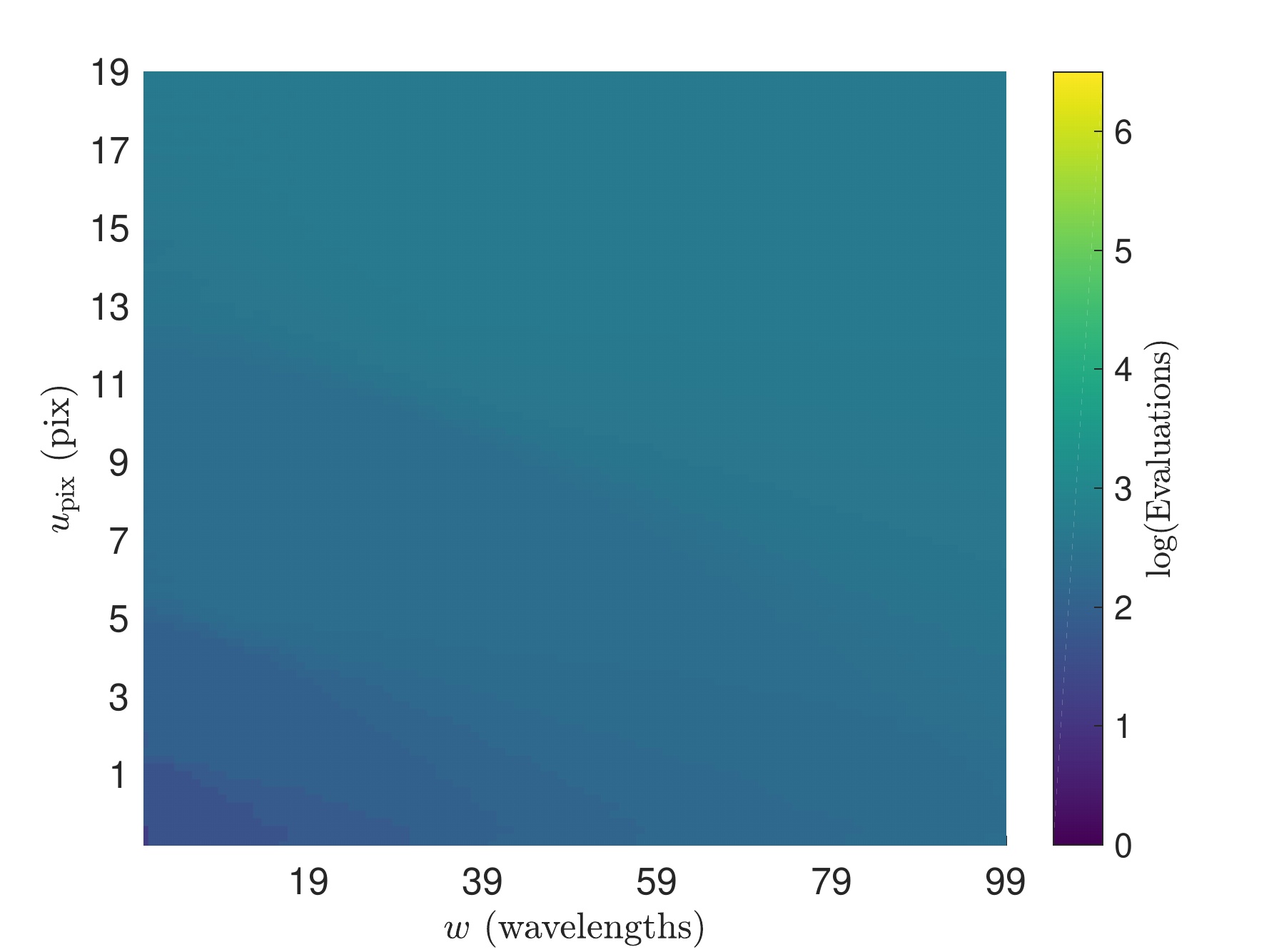}
		\includegraphics[width=6.2cm]{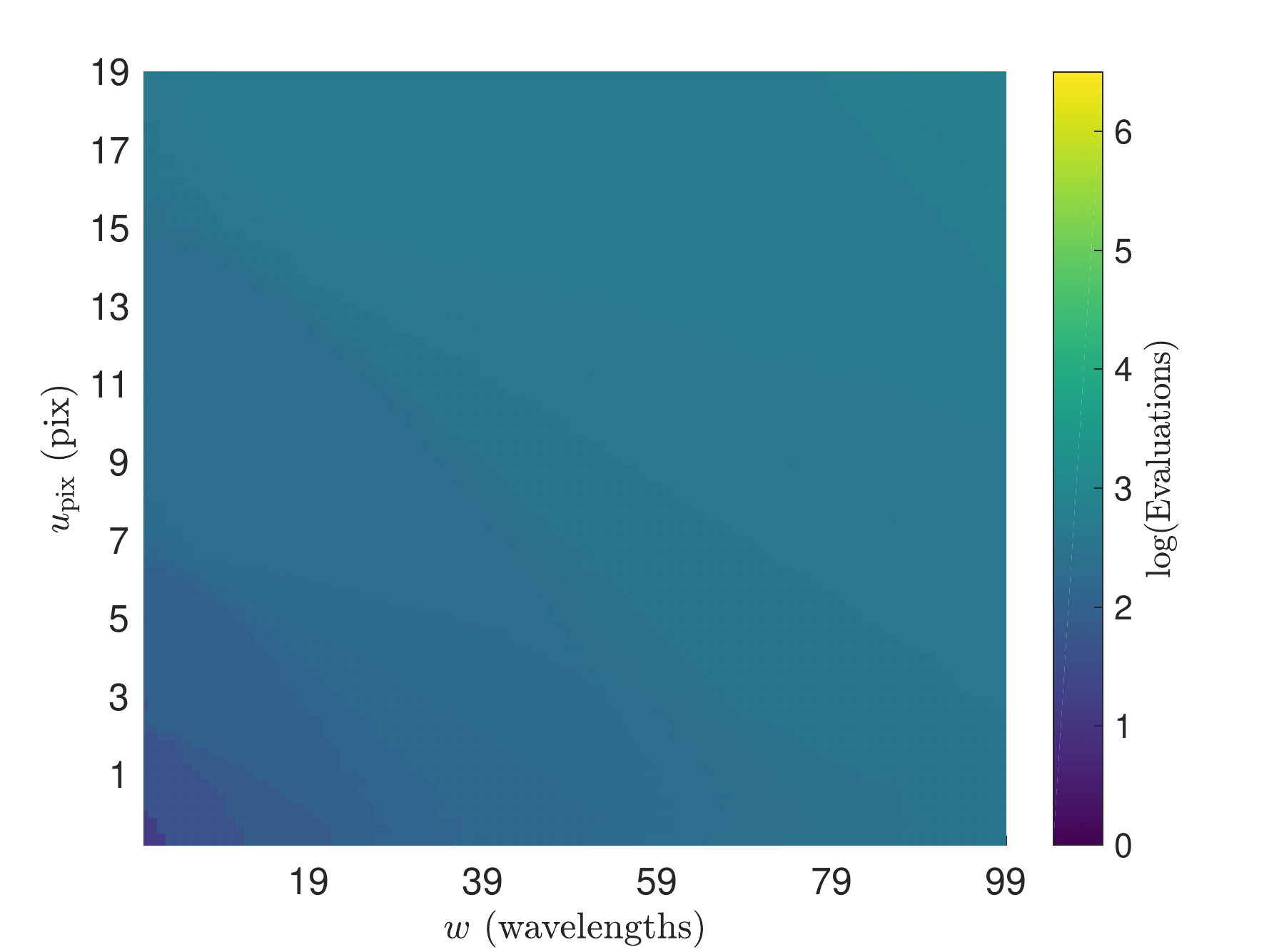}
		\includegraphics[width=6.2cm]{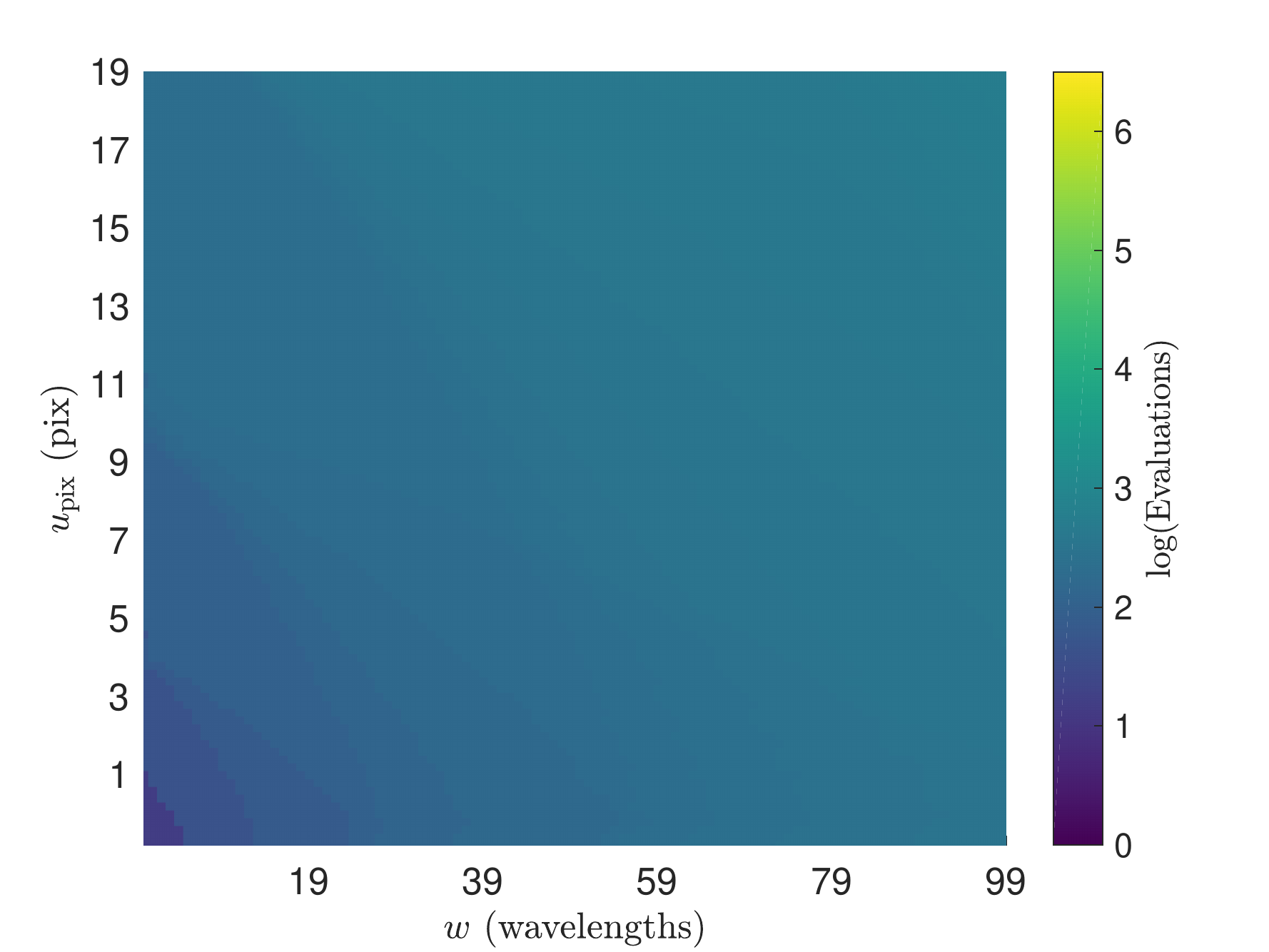}
		\caption{The plots above show the number of function evaluations in the quadrature method required to produce Figures \ref{fig:kernels} (top row) and \ref{fig:kernels_hankel} (bottom row). Each column corresponds to a field of view of {$11.3778^\circ\times11.3778^\circ$}(left), {$17.0667^\circ\times17.0667^\circ$} (middle), and $22.7556^\circ\times22.7556^\circ$ (right). The top row shows two times the values in the bottom row, suggesting that if Equation \ref{eq:analytic_convolution_hankel} takes $N$ evaluations, then Equation \ref{eq:analytic_convolution} takes $N^2$ evaluations to compute. This shows the computation of Equation \ref{eq:analytic_convolution_hankel} scales with radius vs. the computation of Equation \ref{eq:analytic_convolution} that scales with area. The number of evaluations required can be greatly reduced by increasing the absolute error $\eta$.}
		\label{fig:kernels_evals}
	\end{minipage}
\end{figure*}

We then evoke radial symmetry in the gridding kernel and field of view, and evaluate Equation \ref{eq:analytic_convolution_hankel} in Figure \ref{fig:kernels_hankel}. We find that the features of the radially symmetric gridding kernel from Equation \ref{eq:analytic_convolution} match the cross section of Equation \ref{eq:analytic_convolution_hankel}, suggesting little difference between the two kernels. Additionally, when $N$ samples are required to evaluate the 1d radially symmetric kernel, approximately $N^2$ are required to evaluate the 2d kernel, as shown in Figure \ref{fig:kernels_evals}. This suggests that the symmetric kernel calculation scales with radius, not total area as in the 2d case. This has enormous general implications for computation and storage for $w$-projection kernels at large fields of view.
																																								
\subsection{Numerical equivalence of radially symmetric kernel}
Next, we show that using the radially symmetric gridding kernel is consistent with the non radially symmetric kernel.
To test this, we constructed three measurement operators $\bm{\mathsf{\Phi}}_{\rm standard}$ (standard $w$-projection kernel), $\bm{\mathsf{\Phi}}_{\rm radial}$ (symmetric $w$-projection kernel), and $\bm{\mathsf{\Phi}}_{\rm no-projection}$ (no $w$-term), and show that \mbox{$\bm{\mathsf{\Phi}}_{\rm standard} \approx \bm{\mathsf{\Phi}}_{\rm radial}$} within some error (suggesting that they agree), and use $\bm{\mathsf{\Phi}}_{\rm no-projection}$ as a reference operator.
																											
To show that two operators are equivalent, we need the notion of an operator norm $\| \cdot \|_{\rm op}$. The operator norm for an operator that maps between Hilbert spaces ($\ell_2$) has the property that
\begin{equation}
	\|\bm{\mathsf{\Phi}} \bm{x}\|_{\ell_2} \leq \|\bm{\mathsf{\Phi}}\|_{\rm op} \| \bm{x}\|_{\ell_2} \, \quad \forall \bm{x} \in \mathbb{R}^N \, .
\end{equation}
$\|\bm{\mathsf{\Phi}}\|_{\rm op}$ is the smallest value for which this is true for all $\bm{x}$. This allows us to put bounds on the output of $\|\bm{\mathsf{\Phi}}\|_{\rm op}$ for each input. We also have the properties that $\|\bm{\mathsf{\Phi}}\|_{\rm op} = \|\bm{\mathsf{\Phi}}^\dagger\|_{\rm op}$ and \mbox{$\|\bm{\mathsf{\Phi}}^\dagger\bm{\mathsf{\Phi}}\|_{\rm op} = \|\bm{\mathsf{\Phi}}\|_{\rm op}^2$}.
																											
The operator norm allows the following statement
\begin{equation}
\begin{split}
	\frac{\|(\bm{\mathsf{\Phi}}_{\rm standard}-\bm{\mathsf{\Phi}}_{\rm radial})\bm{x}\|_{\ell_2}}{\|\bm{x}\|_{\ell_2}} \quad\quad\quad\quad\\
	\quad\quad\leq \|\bm{\mathsf{\Phi}}_{\rm standard}-\bm{\mathsf{\Phi}}_{\rm radial}\|_{\rm op} \, \quad \forall \bm{x} \in \mathbb{R}^N .
	\end{split}
\end{equation}
 For every input sky model $\bm{x}$, the root-mean-squared (RMS) difference between the model visibilities is bounded by the product of the RMS of the input sky model and the operator norm \mbox{$\|\bm{\mathsf{\Phi}}_{\rm standard}-\bm{\mathsf{\Phi}}_{\rm radial}\|_{\rm op}$}. Additionally, for visibilities $\bm{y}$
\begin{equation}
\begin{split}
	\frac{\|(\bm{\mathsf{\Phi}}_{\rm standard}^\dagger-\bm{\mathsf{\Phi}}_{\rm radial}^\dagger)\bm{y}\|_{\ell_2}}{\| \bm{y}\|_{\ell_2}} \quad\quad\quad\quad\quad\\
	\quad\quad\leq \|\bm{\mathsf{\Phi}}_{\rm standard}-\bm{\mathsf{\Phi}}_{\rm radial}\|_{\rm op} \, \quad \forall \bm{y} \in \mathbb{R}^M .
	\end{split}
\end{equation}
This statement says that the RMS difference between dirty maps is bounded by the product of the RMS of the input visibilities and the operator norm \mbox{$\|\bm{\mathsf{\Phi}}_{\rm standard}-\bm{\mathsf{\Phi}}_{\rm radial}\|_{\rm op}$}. When \mbox{$\|\bm{\mathsf{\Phi}}_{\rm standard}-\bm{\mathsf{\Phi}}_{\rm radial}\|_{\rm op} = 0$}, the two operators will clearly be the same.

Since our linear operators map between two Hilbert spaces, the operator norm of $\bm{\mathsf{\Phi}}$ is the square root of the largest Eigenvalue of $\bm{\mathsf{\Phi}}^\dagger \bm{\mathsf{\Phi}}$. To calculate the largest Eigenvalue, we use the power method (as used in \citet{LP18}).
																											
First we normalize each operator, such that \mbox{$\|\bm{\mathsf{\Phi}}\| = 1$}, so there is no arbitrary scaling. Then we calculate $\|\bm{\mathsf{\Phi}}_{\rm standard}-\bm{\mathsf{\Phi}}_{\rm radial}\|_{\rm op}$ and $\|\bm{\mathsf{\Phi}}_{\rm standard} - \bm{\mathsf{\Phi}}_{\rm no-projection} \|_{\rm op}$.
																											
To construct the measurement operators, we use a variable Gaussian sampling density in $(u, v, w)$, with a root-mean-squared spread of 100 wavelengths. We scale $w$ to have an RMS value of 20 wavelengths. We choose a cell size of 240 arcseconds and an image size of 256 by 256 pixels. This provides a full width field of view of {$17.0667^\circ\times17.0667^\circ$}. It is important to note that the $w$-kernels are a function of the field of view, and not the cell size. The kernel support size is estimated by the $w$-value for each measurement to be $\min(\max(4, 2 w/\Delta u), 40)$. This support has a minimum size of 4 and a largest size of 40, and in between a size of $2w/\Delta u$. The benchmarking was performed on a high performance workstation comprised of two Intel Xeon Processors (E5-2650Lv3) with 12 cores each with 2 times hyper-threading per core (at 1.8 GHz) and 256 Gigabytes of DDR4 RAM (at 2133 MHz).

We found the construction time of a radially symmetric kernel was almost two orders of magnitude faster to calculate. An absolute difference of $10^{-4}$ was used for quantifying quadrature convergence. The power method was considered converged with a relative difference of $10^{-6}$.
																											
\begin{figure}
	\includegraphics[width=8cm]{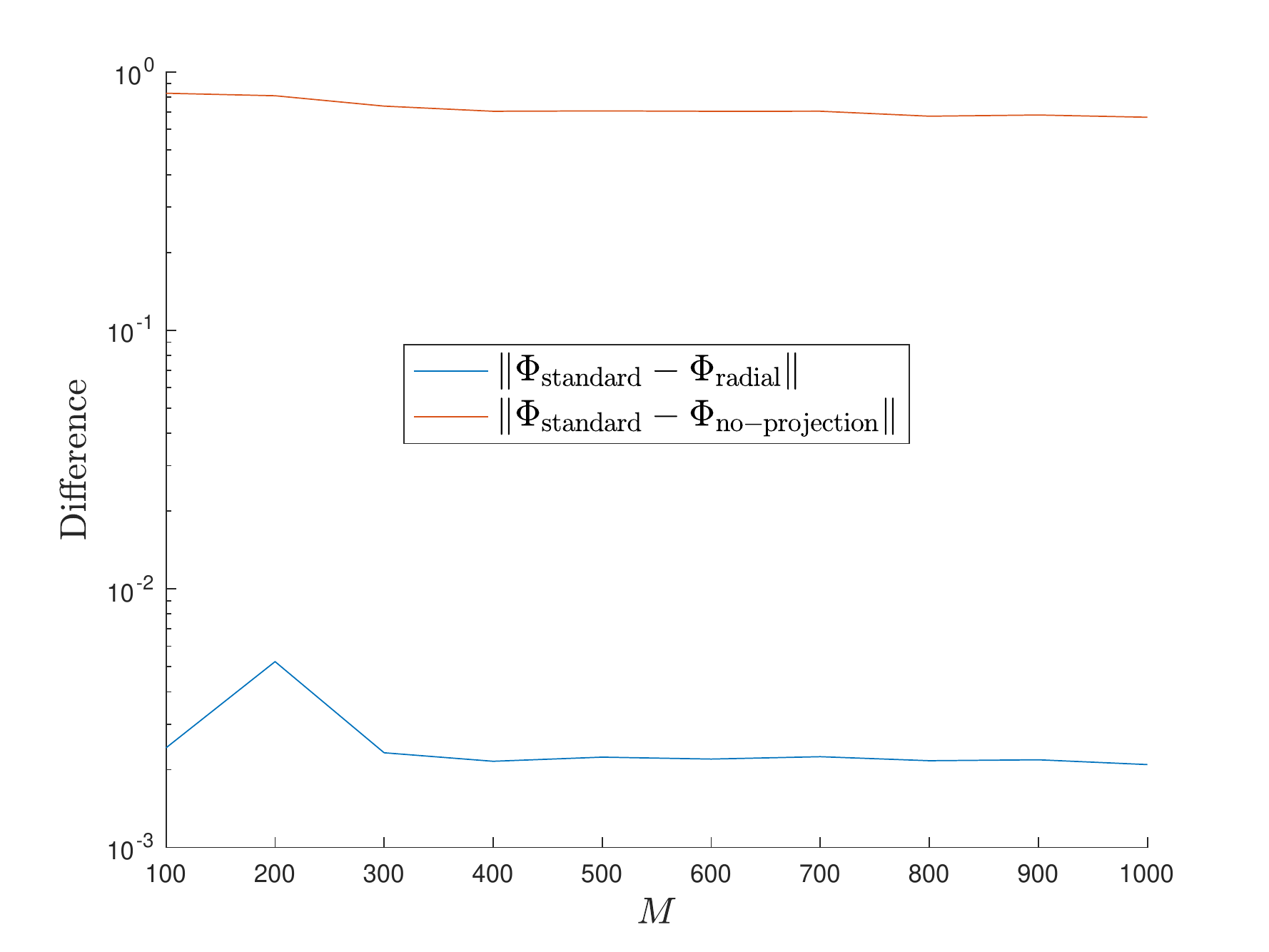}
	\includegraphics[width=8cm]{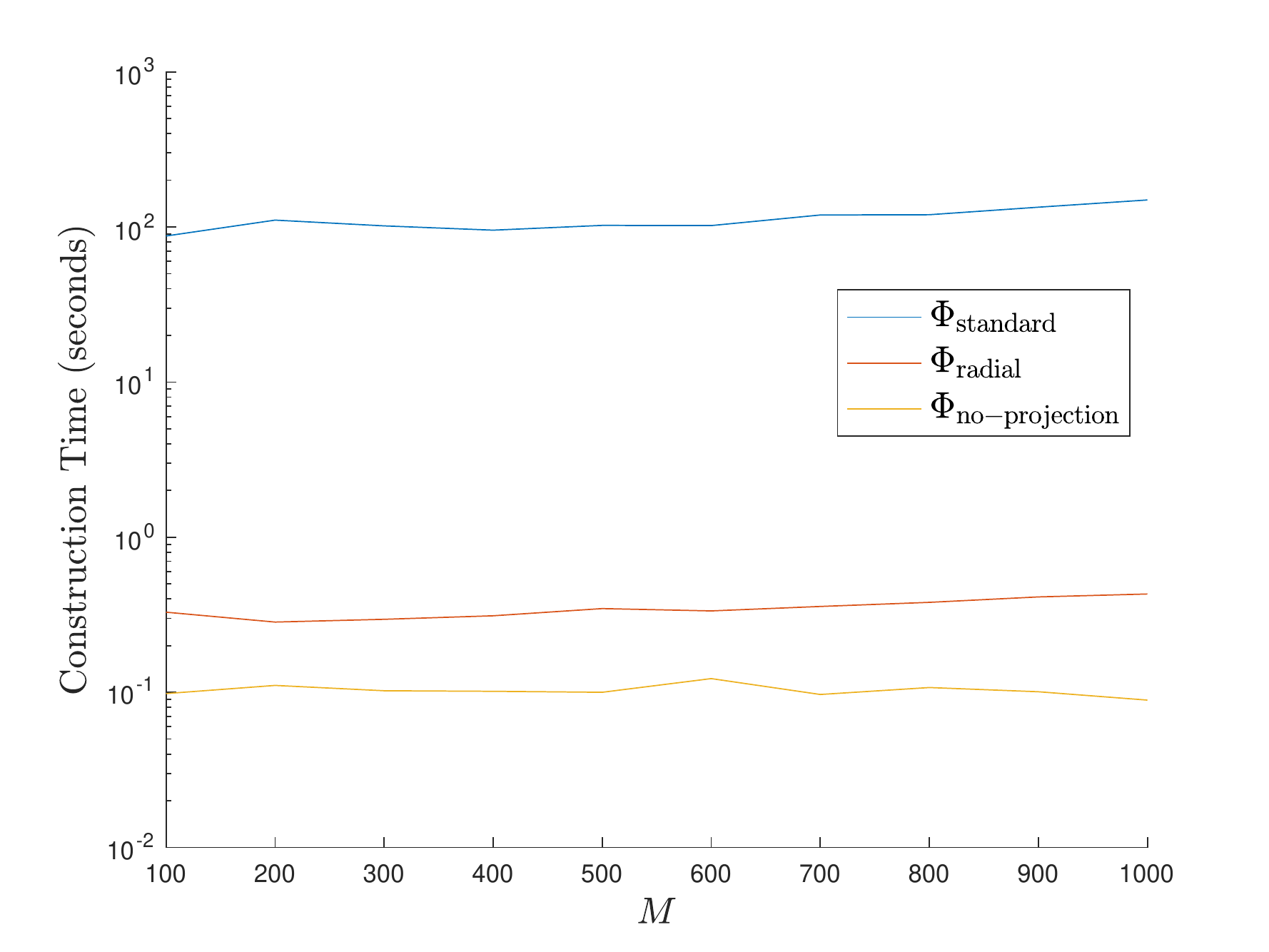}
	\caption{Figures comparing 3 types of measurement operators. One with a standard 2d $w$-projection kernel $\bm{\mathsf{\Phi}}_{\rm standard}$, a radially symmetric kernel $\bm{\mathsf{\Phi}}_{\rm radial}$, and one with no $w$-projection kernel $\bm{\mathsf{\Phi}}_{\rm no-projection}$. The comparisons were performed for 100 to 1000 measurements. (top) The difference in operator norms. We find that the full 2d and radially symmetric kernels are bounded to be the same within about $3 \times 10^{-3}$. We find that assuming no $w$-projection kernel produces a difference close to 1. (Bottom) A plot of the construction time for each operator (excluding normalization). We find that using an analytic expression for the Kaiser-Bessel with no $w$-projection, $\bm{\mathsf{\Phi}}_{\rm no-projection}$, is fastest for two reasons. These are no quadrature integral to calculate, and minimal amount of coefficients to store into memory. The quadrature calculation with variable kernel size means that $\bm{\mathsf{\Phi}}_{\rm radial}$ will always take more time to calculate, even for $w = 0$, which is computationally cheap for quadrature (see Figure \ref{fig:kernels_evals}). We find $\bm{\mathsf{\Phi}}_{\rm standard}$ is the most expensive in time to calculate. This is consistent with the number of function evaluations required to calculate each coefficient.}
	\label{fig:comparison}
\end{figure}

In Figure \ref{fig:comparison}, we show the operator construction time (excluding the normalization), and the operator norm of the difference. Each data point was generated by averaging over 5 realizations. The number of measurements $M$ ranges from only 100 to 1000. From this figure, it is clear that that the operator difference is consistently on the order of $10^{-3}$, suggesting that we have the bounds of $\frac{\|(\bm{\mathsf{\Phi}}_{\rm standard}^\dagger-\bm{\mathsf{\Phi}}_{\rm radial}^\dagger)\bm{y}\|_{\ell_2}}{\| \bm{y}\|_{\ell_2}} \leq 10^{-3} $, which translates to an upper bound dirty map RMS difference of the order of less than 1\%. However, the difference will in principle be less. Similar can be said for generating model visibilities.
																											
It is also clear that the construction times are dramatically different between the two. The construction time is greatly improved by the threading, since the kernel construction was performed in parallel. However, due to the small value of $M$, this improvement has reached saturation. It is clear in this example that construction is hundreds of times faster when using a radial symmetric kernel.
																											
\subsection{Imaging of the directionally dependent $w$-effect via the zero-spacing}
The previous tests have indirectly verified that the radially symmetric $w$-projection kernel is consistent with the 2d $w$-projection kernel, suggesting that the entire degridding and gridding process is self consistent. In this section, we image the generated radially symmetric kernels directly and compare against the theoretically expected values that are independent of implementation. 

In the image domain, we expect the $w$-projection kernel to be a chirp with the form of
\begin{equation}
		c(l, m; w) = {\rm e}^{-2\pi i w(\sqrt{1 - l^2 - m^2} - 1)}\, , 
\end{equation}
then by only imaging the zero-spacing with an artificial $w$-component, which can be done by choosing $y(0, 0, w) = 1$ and $\bar{w} = 0$ in the measurement equation, we find that the adjoint application of the measurement operator and then taking the complex conjugate will result in 
\begin{equation}
dde_{\rm expected}(l, m; w) = a(l, m)\frac{c(l, m; w)}{\sqrt{1 -l^2 - m^2}}\, .	
\end{equation}
It follows that in the discrete setting, gridding a visibility at $(u, v) = (0, 0)$ and $\bar{w} = 0$ will produce the same result
\begin{equation}
	dde_{\rm calculated}(l_i, m_i; w) = \sqrt{N}(\mathsf{\bm{\Phi}}^\dagger_{(u = 0, v = 0, w)})_i^* \, .
\end{equation}

We calculate the average relative difference of $dde$ for the imaginary and real parts, using the formula
\begin{equation}
	\delta (q, p) = 2\left[\frac{q - p}{|q| + |p|}\right]\, ,
\end{equation}
this suppresses divergences for when $q$ or $p$ are close to zero. We choose $a(l, m) = 1$, and values of $w = 10, 100$ wavelengths using an image with 4096 by 4069 pixels and a pixel height and width of 15 arcseconds. This leads to a field of view of {$17.0667^\circ\times17.0667^\circ$}. We compare using a support size linear in $w$, $\frac{2w}{\Delta u}$, rounded to the nearest pixel. We choose an accuracy of $10^{-6}$ in absolute and relative error for numerical quadrature. 

\begin{figure*}
	\begin{minipage}{1.0\textwidth}
		\center
		\includegraphics[width=16cm]{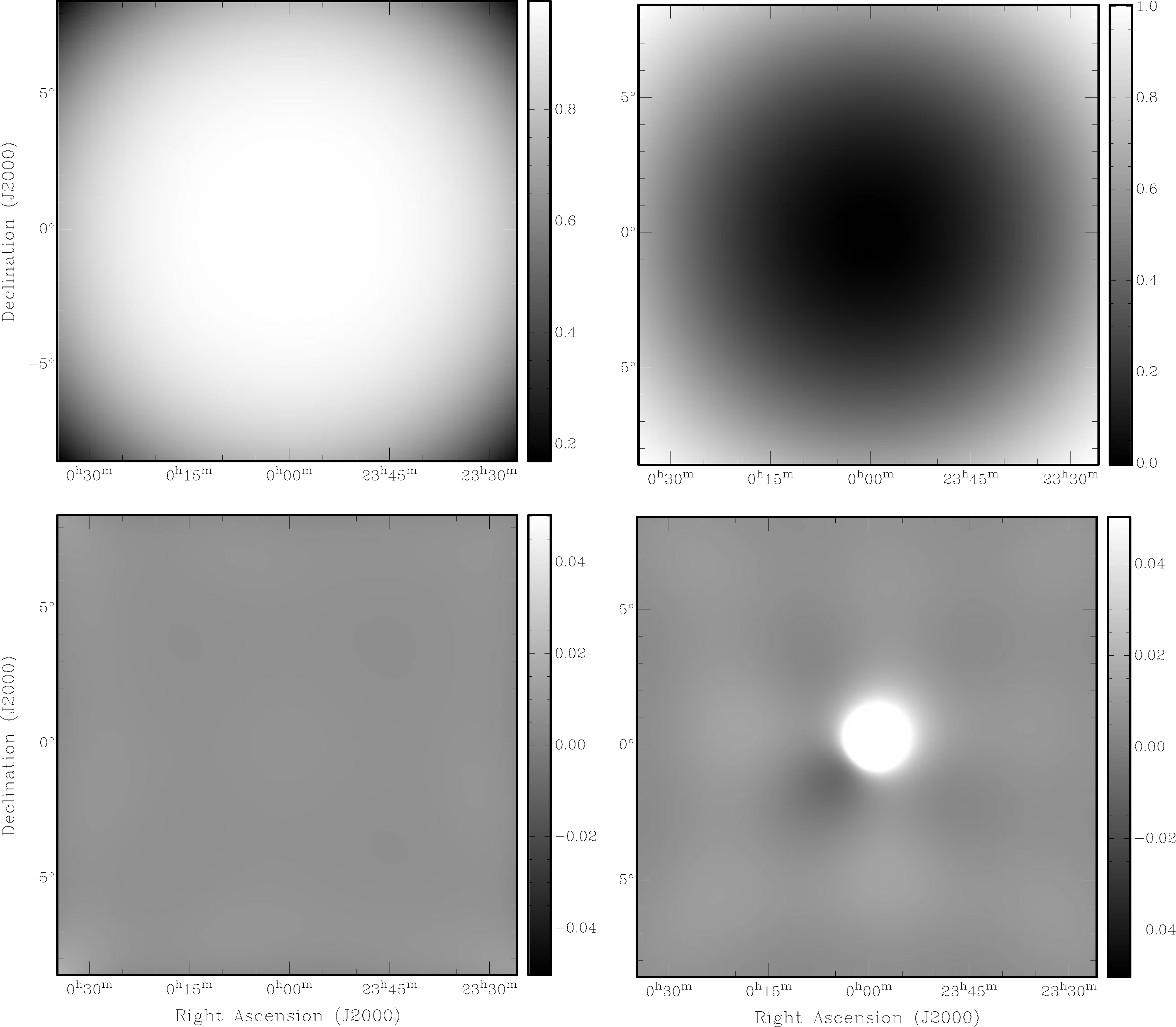}
		\caption{Here we show the calculated radial $w$-projection chirp in the image domain along with the average relative difference of the expected and calculated chirp for both the real and imaginary parts. The left column displays the real component of the chirp, and the right column the imaginary component. The top row is the radial $w$-projection chirp in the image domain calculated using $dde_{\rm calculated}$ with 4096 pixels and a pixel size of 15 arcseconds, calculated for a $w = 10$ wavelengths using a kernel support size of 10 by 10 pixels. The bottom row is the average relative difference $\delta (dde_{\rm expected}, dde_{\rm calculated})$. We find that average relative difference is on the order of 1\%, excluding where $dde_{\rm calculated}$ and $dde_{\rm expected}$ are close to zero and the average relative difference diverges. This shows that the radial symmetric $w$-projection kernel accurately models the directionally dependent $w$-effect at high resolution over wide-fields of view.}
		\label{fig:zero_spacing_10}
	\end{minipage}
\end{figure*}

\begin{figure*}
	\begin{minipage}{1.0\textwidth}
		\center
		\includegraphics[width=16cm]{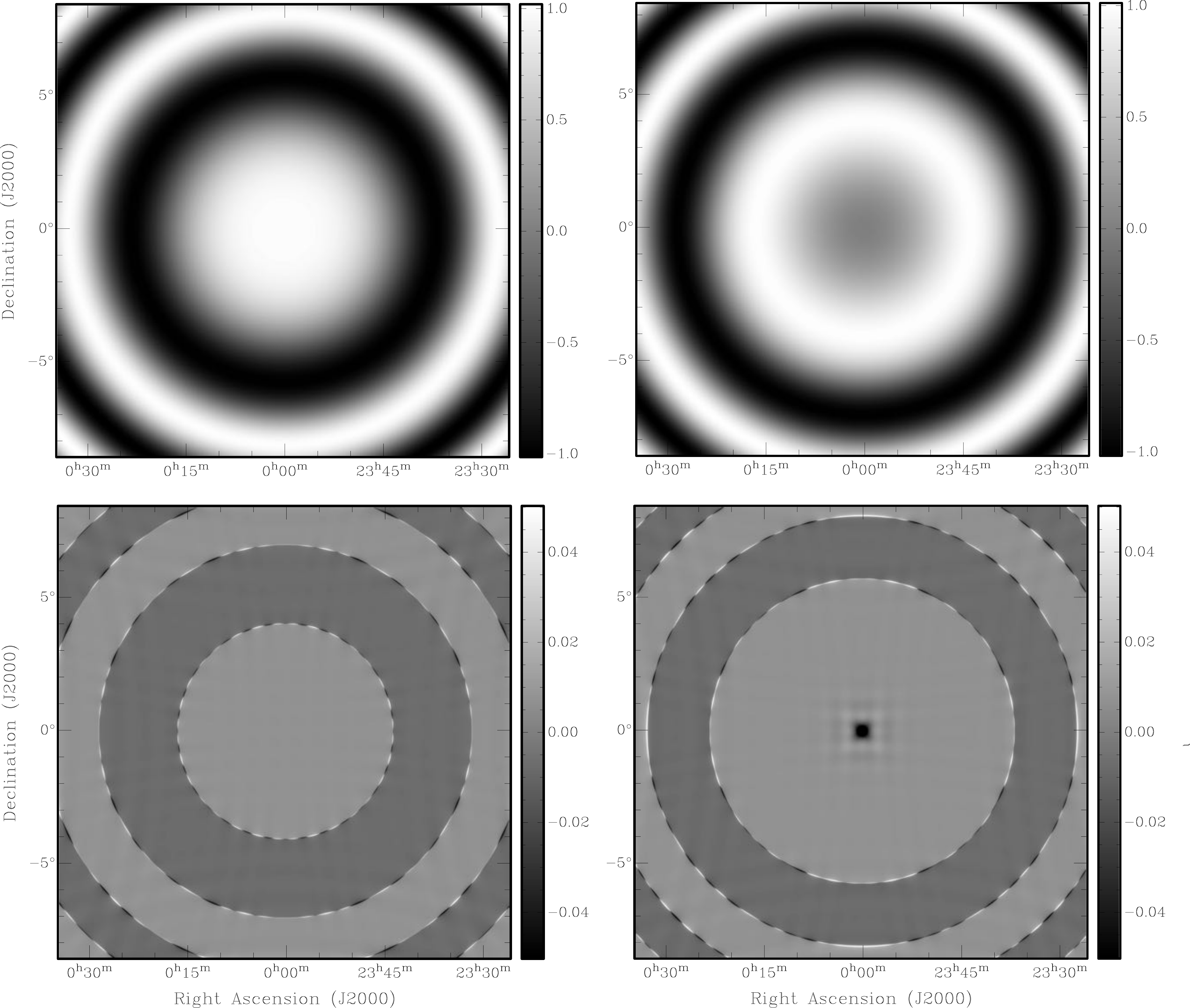}
		\caption{As in Figure \ref{fig:zero_spacing_10}, but for $w = 100$ wavelengths and using a kernel support size of 118 by 118 pixels. Again we find that average relative difference is on the order of 1\%, demonstrating that even for larger $w$, the radial symmetric $w$-projection kernel accurately models the directionally dependent $w$-effect at high resolution over wide-fields of view.}
		\label{fig:zero_spacing_100}
	\end{minipage}
\end{figure*}

Figure \ref{fig:zero_spacing_10} and \ref{fig:zero_spacing_100} show that the radially symmetric $w$-projection kernel has an error on the order of 1\% for both the real and imaginary parts. Where the $w$-effect goes through zero in the real and imaginary parts the average relative difference diverges. It is clear that the $w$-projection kernel still matches the expected $w$-effect, and that these divergences are due to instabilities of the average relative difference for values close to zero.

We find that increasing the support size and reducing the error in numerical quadrature can reduce the average relative difference. We also find that the support size $\frac{2w}{\Delta u}$ and accuracy of $10^{-6}$ in absolute and relative error for numerical quadrature is sufficient for relative error on the order of 1\%. However, if we do not require this accuracy, we can reduce the needed computation by reducing the support size and reducing the accuracy of the numerical quadrature.

{\section{Distributed $w$-stacking $w$-projection hybrid algorithm}
\label{sec:MPI_algo}
In this section, we provide a brief demonstration of using radially symmetric $w$-projection kernels in image reconstruction. We show for the first time that fast and accurate kernel construction, in conjunction with $w$-stacking, enables the ability for modeling sky curvature and non-coplanar baselines to extremely wide-fields of view for each visibility. The kernels are calculated to an absolute accuracy of $10^{-6}$, making the kernel extremely accurate for each $w$ and very wide-fields of view. We present a hybrid of $w$-stacking and $w$-projection algorithm that uses the Message Passing Interface (MPI) standard and show its application to image reconstruction of an MWA observation of Puppis A and Vela. This algorithm is made practical with the developments of the previous section and the use of distributed computation.

\subsection{$w$-stacking-$w$-projection measurement operator}														
First, we distribute the measurements into $w$-stacks using MPI. Then, we generate a $w$-projection kernel for each visibility in a $w$-stack.
																											
The measurement operator corrects for the average $w$-value in the $w$-stack, then applies a further correction to each visibility with the $w$-projection. Each $w$-stack $\bm{y}_k$ has the measurement operator of
\begin{equation}
	\bm{\mathsf{\Phi}}_k = \bm{\mathsf{W}}_k\bm{\mathsf{GC}}_k\bm{\mathsf{F}}\bm{\mathsf{Z}}\bm{\mathsf{\tilde{S}}}_k\, .
\end{equation}
The gridding correction has been modified to correct for the $w$-stack dependent effects, such as the average $\bar{w}_k$ and $1/n(\bm{l})$
\begin{equation}
	\left[\bm{\mathsf{\tilde{S}}}_k\right]_{ii} = \frac{a_k(l_i, m_i){\rm e}^{-2\pi i \bar{w}_k(\sqrt{1 -l^2_i -m^2_i} - 1)}}{g(l^2_i + m^2_i)\sqrt{1 -l^2_i - m^2_i}}\, .
\end{equation}
We choose no primary beam effects within the stack $a_k(l_i, m_i)$.
This gridding correction shifts the relative $w$ value in the stack. This can reduce the effective $w$ value in the stack, especially when the stack is close to the mean $\bar{w}_k$, i.e. to the value of $w_i - \bar{w}_k$\footnote{Another good choice may be to minimize the median $w$ in a stack rather than the mean $w$ in a stack.}. This reduces the size of the support needed in the $w$-projection gridding kernel for each stack,
\begin{equation}
	\begin{split}
		\left[\bm{\mathsf{GC}}_k\right]_{ij} =
		[GC](\sqrt{(u_i/\Delta u - q_{u, j})^2 + (v_i/\Delta u - q_{v, j})^2}\\, w_i - \bar{w}_k, \Delta u)\, .
	\end{split}
\end{equation}
$(q_{u, j}, q_{v, j})$ represents the nearest grid points. For each stack \mbox{$\bm{y}_k \in \mathbb{C}^{M_k}$} we have the measurement equation $\bm{y}_k = \bm{\mathsf{\Phi}}_k\bm{x}$.
																											
To cluster the visibilities into $w$-stacks, it is ideal to minimize the kernel sizes across all stacks, minimizing the memory and computation costs of the kernel. A $k$-means clustering can be used, which greatly improves performance by reducing the values of $|w_i - \bar{w}_k|^2$ across the $w$-stacks. 
																											
It is clear that each stack has an independent measurement equation. However, the full measurement operator is related to the stacks in the adjoint operators such that
\begin{equation}
	\bm{x}_{\rm dirty} =  \begin{bmatrix}\bm{\mathsf{\Phi}}_1^\dagger,& \dots, &\bm{\mathsf{\Phi}}_{k_{\rm max}}^\dagger \end{bmatrix} \begin{bmatrix} \bm{y}_1 \\ \vdots \\ \bm{y}_{k_{\rm max}}  \end{bmatrix} = \bm{\mathsf{\Phi}}^\dagger \bm{y}\, .
\end{equation}
When applying the $w$-stacks in parallel, an MPI all reduce can be used to sum over the dirty maps generated from each node. The full operator $\bm{\mathsf{\Phi}}$ can be normalized using the power method.

\subsection{Distributed Image Reconstruction}																	
For image reconstruction, we use alternating direction method of multipliers as implemented in PURIFY (ADMM) \citep{LP18}, but built using MPI to operate on a computing cluster. The algorithm solves the same minimisation problem stated in \cite{LP18}
\begin{equation}
	\label{eq:l1_analysis}
	\min_{{\bm{x}} \in \mathbb{R}^N}\big\| \bm{\mathsf{\Psi}}^\dagger {\bm{x}} \big\|_{\ell_1}\quad {\rm subject}\, {\rm to} \quad \left \|\bm{y} - \bm{\mathsf{\Phi}} {\bm{x}} \right\|_{\ell_2} \leq \epsilon\, .
\end{equation}
The term $\big\| \bm{\mathsf{\Psi}}^\dagger {\bm{x}} \big\|_{\ell_1}$ is a penalty on the number of non-zero wavelet coefficients, while $\left \|\bm{y} - \bm{\mathsf{\Phi}} {\bm{x}} \right\|_{\ell_2} \leq \epsilon$ is the condition that the measurements fit within a Gaussian error bound $\epsilon$. The wavelet operator $\bm{\mathsf{\Psi}}$ uses a wavelet dictionary of 9 wavelets, which includes a Dirac basis, and Debauches 1 to 8. Each basis in the dictionary $\bm{\mathsf{\Psi}}_k$  has its own node, and is performed in parallel. Like with the adjoint measurement operator, an MPI reduction is performed to sum over the nodes for the forward wavelet operator\footnote{We use the convention that $\bm{x} = \bm{\mathsf{\Psi}} \bm{\alpha}$ and $\bm{\mathsf{\Psi}}^\dagger\bm{x} =  \bm{\alpha}$.}
\begin{equation}
	\bm{x} =  \begin{bmatrix}\bm{\mathsf{\Psi}}_1,& \dots, &\bm{\mathsf{\Psi}}_{9} \end{bmatrix} \begin{bmatrix} \bm{\alpha}_1 \\ \vdots \\ \bm{\alpha}_{9}  \end{bmatrix} = \bm{\mathsf{\Psi}} \bm{\alpha}\, .
\end{equation}

\subsection{MWA observation of Puppis A and Vela}					
We use PURIFY \citep{LP18} and the MPI $w$-stacking $w$-projection hybrid algorithm to reconstruct an observation of Puppis A performed with the MWA telescope. The observation is from the Phase 1 configuration of the MWA taken on 16 May 2013. The data was collected with XX and YY linear polarizations and has been calibrated and flagged following the standard MWA data reduction process, more details on this process be found in \citet{off14}. The observation is centered at (RA = 08:19:59.99, DEC = -42:45:00), with a 112 second integration, and a central frequency of 149.115 MHz with a bandwidth of 30.720 MHz. Figure \ref{fig:puppis_and_vela_coverage} shows a histogram of the visibilities as a function of $w$, the $w$-coverage of the observation ranges between $\pm 600$ wavelengths. The observation contains on the order of 17 million visibilities, and the XX and YY correlations are combined to generate the Stokes I visibilities.

We use a $k$-means algorithm with MPI to sort and distribute the visibilities into 50 $w$-stacks, spread over 25 nodes (2 processes per node, with 1 process per stack), this sorting took approximately 5 seconds. Most $w$-stacks contain $w$-values between 0 and $\pm 12$ wavelengths, however, some stacks contain $w$-values of up to 22 wavelengths. The reconstructed image was performed over a {$25^\circ$ by $25^\circ$} field of view, using $2048^2$ pixels and a pixel width of $45^{\prime\prime}$. Generating the radial $w$-projection kernels took close to 40 minutes, this generation time can be changed with more or less $w$-stacks. Furthermore, the measurement operator was computed in parallel with over 25 nodes, and used in combination with sparse image reconstruction algorithms used in \citet{LP18}. We used the Galaxy Supercomputer (located in the Pawsey Supercomputing Centre\footnote{https://www.pawsey.org.au/our-systems/}).

This observation contains the Puppis A and Vela supernova remnants, a mix of many bright compact sources and extended structures of the galactic plane. With PURIFY, we use natural weighting, as it provides the best performance in modeling both extended and compact structures. We do not include primary beam corrections when solving for the reconstructed image.

Figure \ref{fig:puppis_and_vela} shows the dirty map, residuals, and the reconstructed image. As described in \citet{LP18}, we do not include the restored map, and the reconstructed image is a sky model that is the equivalent to a CLEAN component model. We also follow \citet{LP18} by using the same wavelet dictionary, and scale the epsilon by 275 because the weights are relative not absolute. We can correct the scale of flux due to the field of view by using the Fourier relation $F(\Delta u u_{\rm pix}, \Delta v v_{\rm pix})$ being paired with $\frac{f(l/\Delta u, m/\Delta v)}{\Delta u \Delta v}$. 

To convert the dirty map and residual map to Jy/Beam, we image the weights of the visibilities to obtain the peak pixel value of the point spread function, the dirty map is then divided by this value to convert from Jy/Pixel to Jy/Beam. We find that the residual map has a RMS value of approximately 190 mJy/Beam, with many of the extended structures removed from the residuals. The large scale structures of Vela are accurately removed, with only a few positive regions in the residuals where the negative side-lobes of Vela are located. This shows that the majority of the large scale structures and more compact detailed sources such as Puppis A are accurately modeled using PURIFY over a 25 by 25 degree field of view. The dynamic range of the reconstruction is 19,850.
}
\begin{figure}
		\center
		\includegraphics[width=8cm]{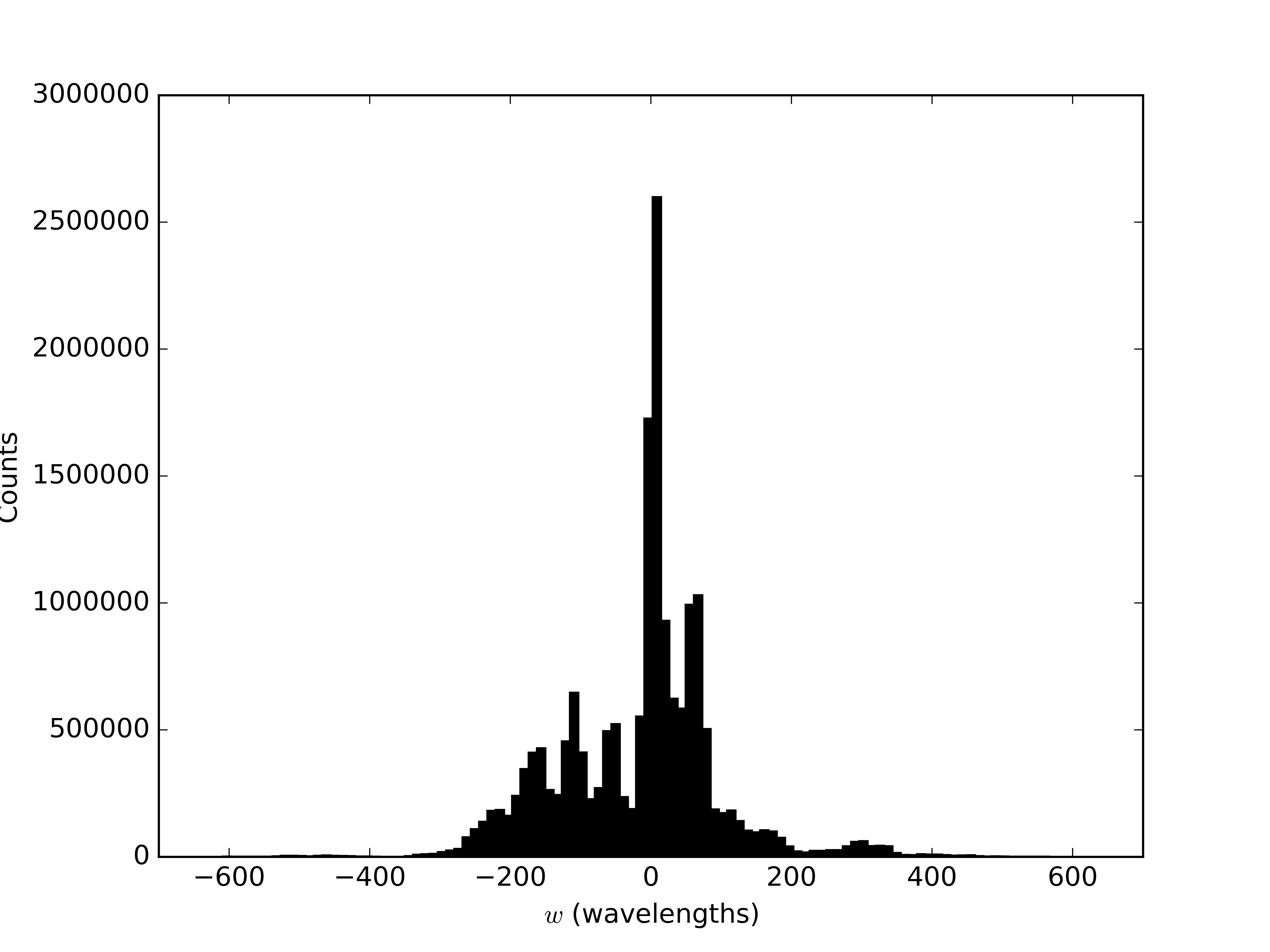}
		\caption{{A histogram of the $w$-coverage of the imaged data using 100 bins. The $w$-values span over $\pm 600$ wavelengths. This $w$-coverage represents 17,529,644 visibilities after flagging of Radio Frequency Interference (RFI) has been applied.}}
		\label{fig:puppis_and_vela_coverage}
\end{figure}			
\begin{figure*}
	\begin{minipage}{1.0\textwidth}
		\center

		\includegraphics[width=16cm]{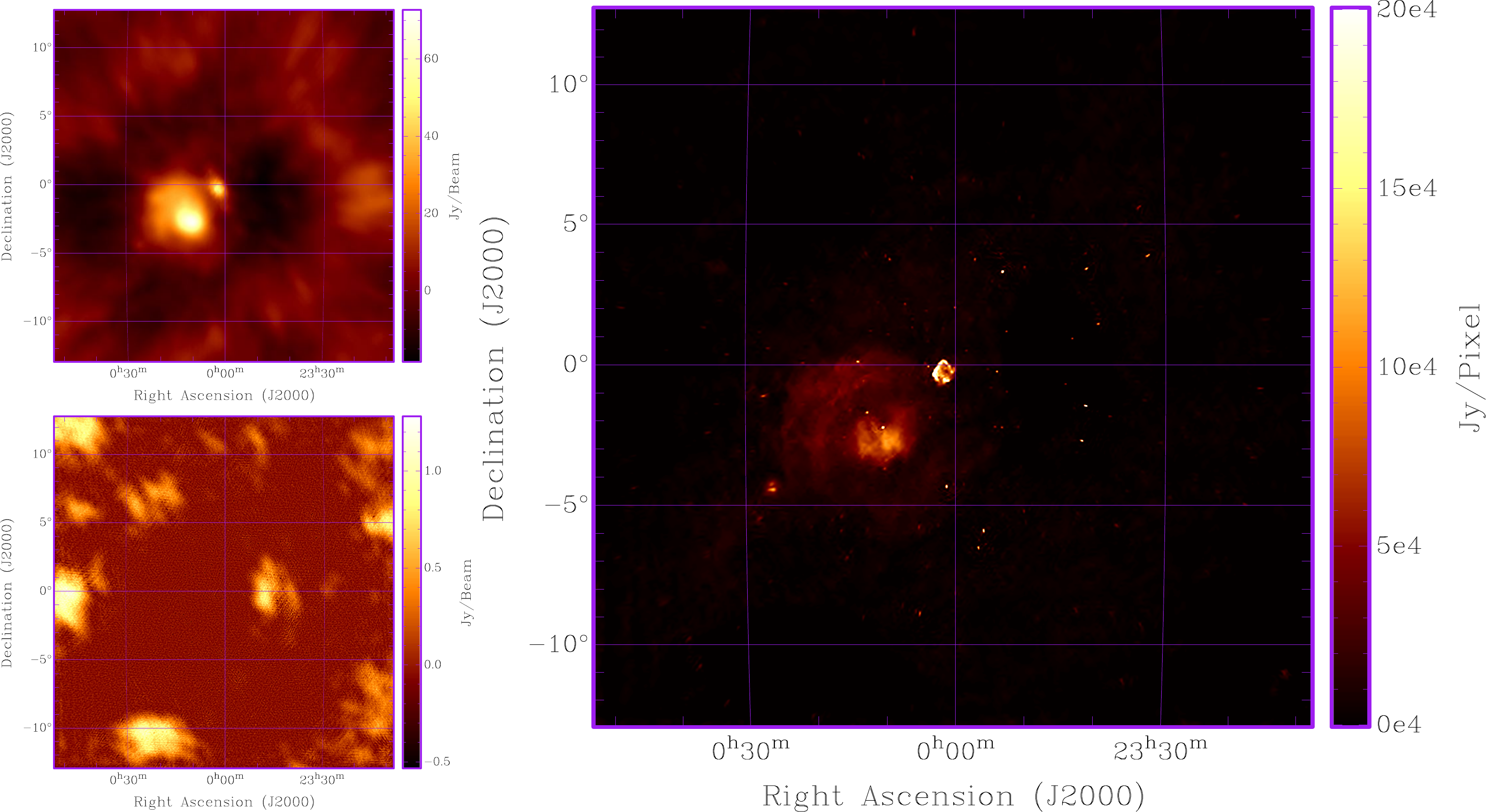}
		\caption{The dirty map (Top Left), residuals (Bottom Left), and sky model reconstruction (Right) of the 112 second MWA Puppis A observation centered at 149.115 MHz, using 17.5 million visibilities and an image size of $2049^2$ (each pixel is 45 arcseconds and the field of view is approximately 25 by 25 degrees). This image was reconstructed using the MPI distributed $w$-stacking-$w$-projection hybrid algorithm, using the radial symmetric $w$-projection kernels, in conjunction with the ADMM algorithm. The RMS of the residuals is 0.189 Jy/Beam, the dynamic range of the reconstruction is 19,850.}
		\label{fig:puppis_and_vela}
	\end{minipage}
\end{figure*}	

\section{Conclusion}
\label{sec:conclusion}
In this work, we investigate exact analytic expressions for modeling curvature in wide-field interferometry, for extremely wide-fields of view. This expression has traditionally been stated in the $(l, m, n)$ domain. However, this work provides the first exact analytic expression for sky curvature and horizon seen in wide-field interferometry in the $(u, v, w)$ domain. Unlike the previous small field of view approximations, this exact kernel does not diverge and is continuous. Furthermore, it provides more insight and understanding of spherical imaging, i.e. it describes a fundamental resolution limit for the measurement of a visibility from a sphere, and the impact of the horizon window in the $(u , v, w)$ domain. While this expression provides insight, the rapid oscillations due to the spherical sky and large support make calculation difficult. These insights suggest that exact computation of projection kernels is more feasible through a Fourier integral from the $(l, m, n)$ domain. 
																																												
As described previously, the effect of the $w$-projection kernel for non-coplanar baselines ($w \neq 0$) becomes greater at larger fields of view. At these extremely wide-fields of view, construction of a $w$-projection kernel is expensive using FFT based methods. Additionally, in this work, we have found that calculations are extremely fast and accurate using adaptive quadrature to compute a radially symmetric gridding kernel. This dramatically reduces the amount of calculations for a numerically exact kernel calculation, reducing the number of samples in the 2d case from $N^2$ to $N$ in the radially symmetric case. 
This immediately makes such a quadrature method computationally competitive. It has low memory usage, it can be distributed in parallel, and scales to extremely wide-fields of view. Furthermore, the calculation is analytic up to a chosen numerical error, allowing the tuning of speed vs. accuracy that is not possible with FFT based methods for large images. 

{In this work, we developed a new technique to validate the calculation and application of a DDE. We show that by applying the modeled DDE when gridding the zero-spacing, we provided an image of the DDE model where it can be directly verified. We applied this to the radial $w$-projection kernel to show the $w$-effect corrections to be accurate on the order of 1\%. This accuracy value is tunable through the support size and the accuracy of the quadrature integration.}
																														
These modeling effects are critical not just for imaging, but calibration of instrumental and ionospheric effects, where the $w$-projection can be used to simulate extremely wide-fields of view. Additionally, any sky model needs to have wide-field of view effects taken into account. Such a sky model maybe critical for physical scientific results. For example, any physical model of the EoR that is to be compared with data collected from a wide-field interferometer needs to have wide-field effects included in the comparison, just as any other instrumental effect (such as the primary beam). This emphasizes that while imaging methods are generally not important for non-imaging experiments, the same process of modeling and correcting for the instrument is still critical in any other analysis.
																																											
The fast and exact correction via quadrature using a radially symmetric kernel is new, and makes fast, exact, spherical and non co-planar baseline corrections possible with a $w$-stacking $w$-projection hybrid. The process works by first correcting for the average $w$-value in a stack to reduce kernel size and total computation, then correcting the exact difference for each visibility using quadrature calculated kernels. {This method was then demonstrated on an MWA observation of the Puppis A and Vela supernova remnants for a 25 by 25 degree field of view and over 17.5 million measurements.}

We have shown that this distributed and paralleled algorithm is extremely powerful for wide-field imaging. Furthermore, these algorithms can be accelerated using multi-threaded parallelism, i.e. General Purpose Graphics Processing Units, in addition to MPI. 
																														
With this work, we provide an important step forward in the fast and accurate evaluation of wide-field interferometric imaging, bringing us closer to solving the computational challenges of the SKA and thus realizing its enormous scientific potential.																																						
% unnumbered section
\section*{Acknowledgements}
LP thanks the Science Technology Facilities Council (STFC) for travel support, and the Curtin University node of International Centre for Radio Astronomy Research (ICRAR) for hosting him during the production of this manuscript.
LP thanks Randal Wayth for his helpful discussion on reading the UVFITS file format. 
This work was supported by the UK Engineering and Physical Sciences Research Council (EPSRC, grants EP/M011089/1) and the UK Science and Technology Facilities Council (STFC, grant ST/M00113X/1). This work was supported by resources provided by The Pawsey Supercomputing Centre with funding from the Australian Government and the Government of Western Australia. We thank the anonymous referee for their help in improving the focus of the manuscript.

{\it Facilities:} MWA, Pawsey Supercomputing Centre																																	
\bibliographystyle{aasjournal}
\bibliography{refs}


\end{document}